%% file: main.tex
\documentclass[11pt]{article}
\usepackage{fancyhdr}
\usepackage{bigints}
\usepackage{mathrsfs}
\usepackage{booktabs}
\usepackage{multirow}
\usepackage{url}
\usepackage[stable]{footmisc}
\usepackage{hyperref}
\usepackage{breakurl}
\usepackage{lineno}
\usepackage{comment}
\usepackage[margin=1in]{geometry}

\usepackage{titlesec}

\usepackage{graphicx} 
\usepackage[thinc]{esdiff}
\usepackage{amsmath}
\usepackage{mathtools}
\DeclarePairedDelimiter\bra{\langle}{\rvert}
\DeclarePairedDelimiter\ket{\lvert}{\rangle}
\DeclarePairedDelimiterX\braket[2]{\langle}{\rangle}{#1\,\delimsize\vert\,\mathopen{}#2}
\usepackage{float}
\usepackage{mwe}
\usepackage{algorithm}
\usepackage{algpseudocode}
\usepackage{xcolor}
\usepackage{subcaption}
\usepackage{soul}
\usepackage[backend=biber,sorting=none]{biblatex}
\input{temp-definitions}
\usepackage{authblk} 
\title{A Variational Quantum Algorithm for Nonlinear Finite Element Analysis of Hyperelastic Materials}

\author{
Uditnarayan Kouskiya$\,^{\ddagger}$,
Caglar Oskay$\,^{\ddagger,\S}$ \thanks{Corresponding author address: VU Station B\#351831, 2301 Vanderbilt Place, Nashville, TN 37235. Email: caglar.oskay@vanderbilt.edu}
}

\date{
$^{\ddagger}$Department of Civil and Environmental Engineering\\
$^{\S}$Department of Mechanical Engineering\\
Vanderbilt University\\
Nashville, TN 37212
}

\begin{document}
\maketitle

\begin{abstract}
\noindent 
This manuscript explores a variational quantum formulation for nonlinear elasticity problems arising from hyperelastic material models. The approach leverages the potential energy structure of hyperelasticity and employs a hybrid quantum–classical framework in which the energy functional is evaluated using parameterized quantum circuits and optimized through classical routines. To enable a hybrid (classical-quantum implementation), polynomial approximations of the nonlinear terms in strain energy density are introduced, yielding a representation compatible with variational quantum algorithms. The methodology is demonstrated on a special case of the Neo-Hookean material model in a one-dimensional setting using finite element discretizations with first- and second-order shape functions and nonhomogeneous boundary conditions. Numerical experiments investigate the influence of the polynomial approximation order on the accuracy and efficiency of the proposed approach, illustrating its feasibility for near-term quantum devices.

\end{abstract}

\section{Introduction}

Numerical evaluation of nonlinear partial differential equations (PDEs) sits at the foundation of computational mechanics. A common approach to solving nonlinear mechanics problems is to obtain an algebraic system following discretization, which is then solved iteratively. The computational cost of the overall process grows rapidly with the system size. As demand for simulation of large, complex material and structural systems increases, such computations place significant pressure on conventional computing architectures, particularly due to performance, memory, and energy-efficiency limitations associated with the von Neumann computing architectures \cite{10.1145/3665283.3665293}. Quantum computing (QC) has recently emerged as an alternative computing paradigm, owing to the exponential growth of its accessible state space with the number of qubits \cite{yang2023survey,childs2023improved, Balducci2022/fmech.2022.914241}. In this work, we take an initial step in developing and investigating a variational quantum algorithm directed toward problems in nonlinear elasticity by considering a special case of the Neo-Hookean material model as a representative case. 

The majority of the relevant existing QC literature focuses on the evaluation of linear ordinary differential equations (ODEs) and PDEs. A common solution approach is to construct a linear system of equations following discretization, and evaluating it using quantum linear systems algorithms (QLSAs)~\cite{berry2017quantumODE, liu2022thesis}. The foundational work by Harrow et al.~\cite{harrow2009quantum} (HHL) introduced the first QLSA, which solves sparse linear systems $Ax=b$ by encoding the right-hand side vector $b$ as a quantum state $|b\rangle$. Using Hamiltonian simulation~\cite{lloyd1996universal} and quantum phase estimation, the algorithm prepares a state $|x\rangle \propto A^{-1}|b\rangle$ that encodes the properties of the solution vector $x$.
Subsequent developments produced high-precision QLSA variants based on linear combination of unitaries (LCU)~\cite{berry2012lcu}, 
quantum singular value transformation (QSVT)~\cite{martyn2021grand}, and adiabatic approaches~\cite{subasi2019quantum}. However, despite their favorable asymptotic complexity, the aforementioned approaches require deep circuits and error rates that are beyond the capabilities of the NISQ-era hardware~\cite{preskill2018quantum}. 

Within the framework of near-term quantum computing, Variational Quantum Algorithms (VQAs) have emerged as a prominent and widely explored class of algorithms \cite{bharatiVQAreview, VQEfirstpaper, cerezo2021vqareview}. The central idea behind VQAs is to offload the computationally intensive parts of a problem to a quantum processor that evaluates a cost function using parameterized quantum circuits encoding the quantity of interest. The circuit parameters are then iteratively updated by a classical optimizer to minimize this cost function. VQAs face well-documented trainability challenges.  
As circuit width or ansatz depth increases, the cost landscape can develop barren plateaus, regions where gradients vanish exponentially and classical optimization stalls~\cite{mcclean2018barren}; this behavior is closely tied to the choice of cost function and the ansatz, with global observables inducing barren plateaus more readily than local ones~\cite{cerezo2021costfunction}. The barren plateau problem is compounded by hardware noise~\cite{wang2021noiseinduced}.

Several efforts have applied general-purpose quantum algorithms to problems in computational mechanics. For example, the Poisson equation and elliptic finite element problems have been addressed via QLSAs~\cite{cao2013poisson,montanaro2016fem}, lying in the fault-tolerant quantum computing regime. Similarly, Liu et al.~\cite{liu2024mechanics} demonstrate an algorithm for the linear elastic homogenization of periodic composites using a quantum Fourier transform (QFT)-based approach. Within the VQA regime, the Variational Quantum Linear Solver (VQLS)~\cite{BravoPrieto2023VQLS} emerged as one of the prominent VQAs for linear systems, and was used by Arora et al.~\cite{Arora:2025a} for FEM-based steady-state heat conduction, and Rao et al.~\cite{rao2024vqlsElasticity} for plane strain elasticity problems. Raisuddin and De~\cite{raisuddin2023feqa} instead employed  quantum annealers to solve linear finite element systems via an equivalent quadratic optimization formulation. 

The extension of quantum algorithms to nonlinear problems in mechanics remains challenging due to the inherently linear structure of quantum operations. Several existing approaches reformulate nonlinearities in the governing equations into representations compatible with available quantum algorithmic primitives, differing primarily in the quantities being expanded and the resulting reformulation. Kyriienko et al.~\cite{kyriienko2021dqc} represent the solution function through a truncated Chebyshev quantum feature map, while the nonlinear operations associated with the governing equation are evaluated through classical post-processing.  Xu et al.~\cite{xu2024anm} expand the solution path with respect to a continuation parameter, resulting in a sequence of linear subproblems solved using VQLS. Endo and Takahashi~\cite{endo2024divergence} employ Carleman linearization to embed a nonlinear differential equation into a larger truncated linear system. Soize et al.~\cite{soize2025fkp} apply a similar strategy to a Fokker–Planck eigenvalue problem, which, although linear, involves an underlying non-algebraic potential approximated via a truncated polynomial expansion to enable a Variational Quantum Eigensolver-based solution.

In the VQA regime, Lubasch et al.~\cite{lubasch_zoo,jaksch2023vqacfd} introduced a novel VQA framework for nonlinear PDEs with polynomial nonlinearities by devising quantum nonlinear processing units and demonstrated the efficacy of this approach on Burgers and nonlinear Schrödinger equations. Sarma et al.~\cite{samra2023vqapde} recently extended this approach to solve a broader class of nonlinear PDEs. The idea expressed in these works relies on a least-squares functional as the cost function. Sato et al.~\cite{PhysRevA.104.052409} proposed a VQA to solve the Poisson equation by formulating the corresponding  Dirichlet energy as a cost function parameterized by a quantum circuit. The energy terms were evaluated using a quantum oracle, and the circuit parameters were optimized classically to minimize the cost, thereby encoding the solution. Similarly, Over and Bengoechea et al. \cite{Over2025Boundary,bengoechea2025quantum}, also derived a variational form for the advection–diffusion equation in the time-discrete setting and minimized it within a VQA framework.

This manuscript presents a VQA-based approach for nonlinear elasticity problems arising from hyperelastic material models under certain assumptions. The proposed framework exploits the fact that, for hyperelastic materials, the stress is derived from a strain energy density function, thereby enabling a variational formulation in terms of a total potential energy functional \cite{marsden1994mathematical,bonet1997nonlinear,ogden1997nonlinear,holzapfel2000nonlinear,ciarlet1988mathematical}. 
 The resulting approximated potential energy functional is constructed to integrate naturally within a variational quantum framework. The proposed methodology is demonstrated on a simplified one-dimensional Neo-Hookean material-based problem. Two distinct approximation strategies are developed and implemented in accordance with contemporary VQA capabilities. The displacement field is discretized using first- and second-order finite element shape functions, and the formulation accommodates nonhomogeneous boundary conditions. The influence of higher-order polynomial approximations on the efficiency and accuracy of the VQA implementation is evaluated. Numerical verification of the proposed approach has been performed using noise-free statevector simulations rather than execution on real quantum hardware. We discuss several practical algorithmic and hardware-based challenges towards a practical realization of the framework and extension to multi-dimensional problems. The  novel contributions of  the proposed approach are therefore as follows: (1)~The numerical problem is expressed as the minimization of the variational form (as opposed to the weak form) to prepare it for VQA; and (2)~treatment of the nonlinear terms in the variational form using the QNPUs.

The remainder of this manuscript is organized as follows. Section~\ref{sec:overview} presents the overview of the proposed approach. The governing equations for a general multidimensional problem are first introduced and then restricted to the one-dimensional setting. The resulting potential energy is then embedded into a VQA framework, concluding with an outline of the overall VQA procedure. Section~\ref{sec:NeoHookean_developement} applies the scheme to a special case of the Neo-Hookean material model, introducing the energy functional, deriving a quantum-computing-compatible approximation, and constructing its discretized form using finite element shape functions and Gaussian quadrature. Section~\ref{sec:VQA_for_neohookean} provides the details of the quantum implementation. The basic quantum circuits used as building blocks, the state preparation procedure for encoding real-valued vectors, and the evaluation of the individual terms in the cost function based on specifically designed quantum circuits are described. In Section~\ref{sec:complexity}, we analyze the gate and time complexity of the proposed approach. Section \ref{sec:examples} presents numerical results demonstrating the performance of the VQA for different polynomial approximation orders and for first- and second-order finite element discretizations. Section \ref{sec:challenges} discusses the practical challenges associated with realizing the proposed framework and outlines potential future directions for addressing some of these issues. Section~\ref{sec:conclude} summarizes the main findings and provides a conclusion.

\section{Overview of the Proposed Approach}
\label{sec:overview}
In this section, we outline the proposed VQA-based approach for nonlinear elasticity problems involving hyperelastic materials. We begin by formulating the governing equations and deriving a simplified model suitable for analysis. The VQA framework requires a cost function to optimize, which we construct from the system’s potential energy. To evaluate this energy on a quantum computer, the displacement field (represented by its discretized values) must be encoded as a quantum state. This motivates the introduction of a quantum representation of real-valued vectors. Finally, we integrate these components to establish a complete framework for approximating the solution to the underlying elasticity problem.

\subsection{Mathematical model} \label{sec:math_model}

The nonlinear elastostatic boundary-value problem in the reference configuration $\bbOmega_{0}\subset\mathbb{R}^{3} (\bbX = (X,Y,Z))$ is given by
\begin{subequations} \label{general_BVP}
\begin{align}
\label{eq:governing_1}
\nabla_{\bbX}\cdot\bbP(\bbF(\bbX)) + \bbB(\bbX) = \mathbf{0}&
\quad \text{for }\bbX \in \bbOmega_{0}, \\ 
\bbF = \nabla_{\bbX} \boldsymbol{\varphi}  =  \bbI + \nabla_{\bbX} \mathbf{u}(\bbX)& \\
\mathbf{u}(\bbX) = \bar{\mathbf{u}}(\bbX) &
\quad\text{for } \bbX \in \Gamma_{D}, \\
\qquad
\bbP \cdot \bbN = \boldsymbol{\bar{P}}(\bbX)&
\quad\text{for } \bbX \in \Gamma_{N},
\end{align}
\end{subequations}
where $\bbP$ is the first Piola-Kirchhoff stress tensor, $\boldsymbol{\varphi}$ the deformation map, $\mathbf{u}=(u,v,w)$ the displacement, $\bbF$ the deformation gradient and $\bbB$ the body force per unit reference volume. For a second-order tensor field $\bbT(\bbX)$, $(\nabla_{\bbX}\cdot\bbT)_a = \partial T_{aA}/\partial X_A$~\cite{holzapfel2000nonlinear,bonet1997nonlinear}.
$\bbN$ is the outward unit normal at the reference domain boundaries, and $\bar{\mathbf{u}}$ and
$\boldsymbol{\bar{P}}$ are prescribed displacements and tractions, respectively. The Dirichlet and Neumann boundaries are denoted as $\Gamma_{D} \subset \partial \bbOmega_{0}$ and $\Gamma_{N} \subset \partial \bbOmega_{0}$, respectively, such that $\Gamma_{D}\cup\Gamma_{N}=\partial\bbOmega_{0}$ and $\Gamma_{D}\cap\Gamma_{N}=\emptyset$.
For hyperelastic materials, the stress follows from a strain energy density
$W(\bbF)$ via $\bbP = \partial W/\partial\bbF$. For such materials, it is a classical result that there exists a potential energy functional
\begin{equation}
\mathcal{E}[\mathbf{u}]
    = \int_{\bbOmega_{0}} W(\bbF)\,\mathrm{d}V
      - \int_{\bbOmega_{0}} \bbB\cdot\mathbf{u}\,\mathrm{d}V
      - \int_{\Gamma_{N}} \boldsymbol{\bar{P}}\cdot\mathbf{u}\,\mathrm{d}A,
      \label{eq:mardsen_energy}
\end{equation}
up to a constant that depends on the applied traction and the body force. The critical points of $\mathcal{E}$ satisfy the weak form of the governing equations \eqref{general_BVP}. The number and nature of the critical points (i.e., minima vs. saddle points) of the potential energy functional depend on the properties of $W$~\cite{marsden1994mathematical}.

We consider a one-dimensional reduction of the above three-dimensional theory. Specifically, we assume that the reference configuration is a prismatic body of constant cross-sectional area $A$, such that
\begin{equation}
\bbOmega_0 = (0,L) \times A,
\end{equation}
where $A \subset \mathbb{R}^2$. We further assume that all fields depend only on the axial coordinate $X$, and that the deformation is uniaxial along the $X$-direction. The lateral surface $(0,L)\times \partial A$ is assumed to be traction-free, while the remainder of the boundary conditions are prescribed on the end faces at $X=0$ and $X=L$.

Under these assumptions, the deformation gradient reduces to
\begin{equation}
\label{eq:def_grad_1}
\bbF(\bbX) = \operatorname{diag}(\lambda(X), 1, 1),
\end{equation}
where
\begin{equation}
\lambda(X) = 1 + u'(X)
\end{equation}
represents the principal stretch in the axial direction and $(\cdot)' \equiv {\mathrm{d}(\cdot)}/{\mathrm{d}X}$. Physical admissibility requires
\begin{equation}
J(\bbX) = \det \bbF(\bbX) = \lambda(X) > 0, \quad \forall \, X \in (0,L)
\end{equation}

Since all fields are uniform over the cross-section, volume and surface integrals factor into the cross-sectional area $A$, multiplied by one-dimensional integrals over $(0,L)$. For simplicity, and without loss of generality, we work per unit reference cross-sectional area (i.e., $|A|=1$), so that the three-dimensional formulation reduces to a one-dimensional problem on the interval
\begin{equation}
\Omega_0 = (0,L),
\qquad
\partial \Omega_0 = \{0,L\}.
\end{equation}
The Dirichlet and Neumann boundaries therefore correspond to subsets of the endpoints.

Consequently, we consider the one-dimensional reduction of the potential energy functional: \eqref{eq:mardsen_energy}:
\begin{equation}
\mathcal{E}[u]
:= \int_{\Omega_0} W\big(\lambda)\,\mathrm{d}X
- \int_{\Omega_0} B(X)\,u(X)\,\mathrm{d}X
- \sum_{X\in\Gamma_N} \bar{P}\,u(X),
\label{eq:mardsen_1d}
\end{equation}
where $B$ represents the body force, $\bar{P}$ the prescribed axial traction and $W$ is a function of $\lambda$ only. 
The stationary point of the above functional yields the following weak form:
\begin{quote}
For $u=\bar{u}$ on $\Gamma_D$, find $u$ such that
\begin{equation}
\int_{\Omega_0} P(\lambda)\, v'\,\mathrm{d}X
=
\int_{\Omega_0} B\, v\,\mathrm{d}X
+ \sum_{X\in\Gamma_N} \bar{P}\, v,
\quad \forall v \in H^1(\Omega_0); \quad v=0 \mbox{ on } \Gamma_D 
\end{equation}
where $P(\lambda) = {\partial W}/{\partial \lambda}$.
\end{quote}

Under sufficient smoothness assumptions on the displacement field, the weak form is equivalent to the strong form (equilibrium equation) in the axial direction, given by:
\begin{equation}
\label{eq:1d_equilibrium}
\frac{\mathrm{d}}{\mathrm{d}X}\big(P(\lambda(X))\big) + B(X) = 0
\quad \text{in } \Omega_0,
\end{equation}
subject to the boundary conditions
\begin{equation}
u = \bar{u} \quad \text{on } \Gamma_D,
\qquad
P = \bar{P} \quad \text{on } \Gamma_N.
\end{equation}

\subsection{Quantum representation of real vectors}
\label{sec:quantum_rep}
Unlike classical computing where information is represented using \emph{bits} that take values of $0$ or $1$, quantum computing uses \emph{qubits}—quantum two-level systems—that can exist in a superposition of the $0$ and $1$ states. The computational basis states for a single qubit are denoted as $\ket{0}$ and $\ket{1}$. For a register of $n$ qubits, the corresponding Hilbert space has dimension $N_q = 2^n$. The computational basis states are denoted as
\begin{equation}
\left\{ \ket{\mathrm{bin}(k)} \ \middle| \ k = 0, 1, \ldots, 2^n-1 \right\},
\end{equation}
where $k$ is the integer label of the basis state, and its binary representation specifies the state of each qubit.  
For example, in a three-qubit system, $\ket{\text{bin}(5)}$ corresponds to the quantum state $\ket{101}$, which indicates that the first and third qubits are in state $\ket{1}$, while the second qubit is in state $\ket{0}$. In what follows, we use $\ket{k}$ instead of $\ket{\text{bin}(k)}$ for simplicity, unless otherwise stated.

Any real vector $\bbp\in\mathbb{R}^{N_q}$ (complex vectors are admissible, but we restrict our attention to real-valued vectors), is encoded as a quantum state $\ket{\hat{p}}$ (or simply $\hat{p}$ when no confusion arises), such that:
\begin{equation}
\label{eq:f_represent}
  \sum_{i=0}^{N_q-1} p_i \,\bbe_i =  \bbp  \;=\; \|\bbp\|\, \ket{\hat{p}} \;=\; \|\bbp\| \sum_{k=0}^{N_q-1} \hat{p}_k \ket{k},
\end{equation}
noting that the quantum state $\ket{\hat{p}}$ is normalized.

\begin{figure}
    \centering
    \includegraphics[width=1\textwidth]{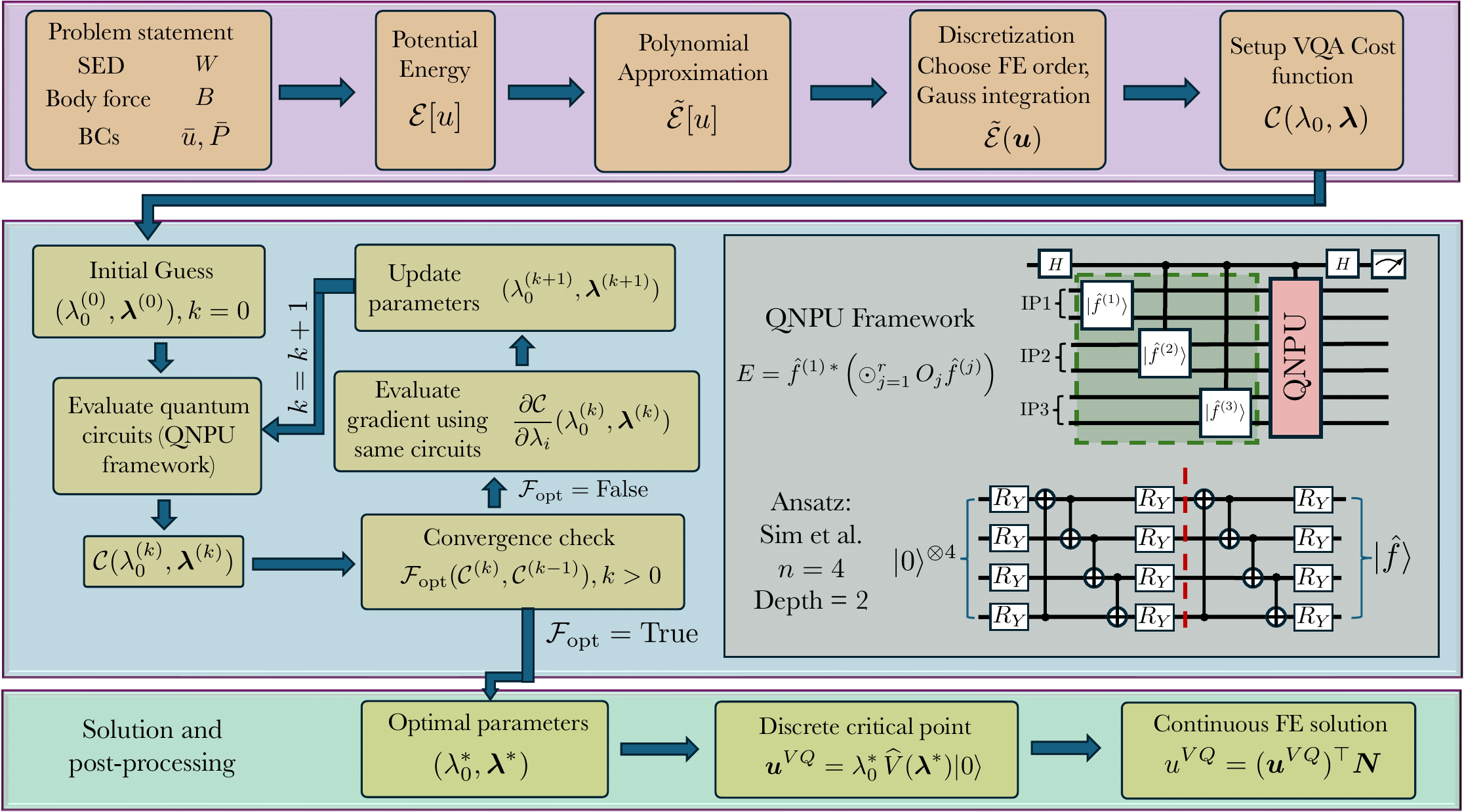}
    \caption{VQA Framework.}
    \label{fig:VQA_outline}
\end{figure}

Any quantum state $\ket{\hat{p}}$ is prepared on a quantum computer using a parameterized quantum circuit, or \emph{Ansatz} circuit, represented by a unitary operator $\widehat{V}(\boldsymbol{\lambda}_p)$, where 
\begin{equation}
\boldsymbol{\lambda}_p = (\lambda_{p_1}, \lambda_{p_2}, \ldots, \lambda_{p_m}); \quad \lambda_{p_i}\in [0,2 \pi],
\end{equation}
is the set of ansatz parameters (each element corresponds to a rotation angle applied to a qubit). When applied to the initial basis state $\ket{0}$, it yields:
\begin{equation}
    \ket{\hat{p}} \;=\; \widehat{V}(\boldsymbol{\lambda}_p) \, \ket{0}.
\end{equation}
The ansatz defines a parameterized trial space of quantum states, whose structure determines its ability to approximate the target state. Different ansatz constructions employ specific arrangements of single-qubit rotation gates and two-qubit entangling gates, which govern the expressiveness of the resulting state space. In this work, we adopt the ansatz proposed by Sim et al.~\cite{sim_et_al}, shown in Fig.~\ref{fig:VQA_outline}. Other ansatz choices with sufficient expressiveness could be used in its place without loss of generality.
Ansatz parameters
along with the scaling parameter $\lambda_{p_0}:=\|\bbp\|$, are collectively referred to as the \emph{control parameters} and generate the vector $\bbp$ based on \eqref{eq:f_represent}.

\subsection{Variational Quantum Algorithm}

The overall structure of the proposed VQA framework is depicted in Fig.~\ref{fig:VQA_outline}. The workflow consists of two main phases: a setup phase (purple) and an optimization phase (blue). At its core, the method represents the unknown displacement field through a set of tunable control parameters, which are iteratively adjusted to approximate the solution.

In the setup phase, the elasticity problem is specified via its strain energy density, which defines the corresponding potential energy functional. This functional is considered for evaluation using quantum circuits, with the control parameters serving as circuit input. Instead of assembling element-wise contributions, the quantum approach directly estimates global or macroscopic quantities (expectation values) of interest. In particular, expectation values from a small set of circuits, are combined classically to estimate the total potential energy of the system.

A key challenge, however, arises from the inherently linear nature of quantum operations, which makes the direct implementation of nonlinear transformations involved in the potential energy difficult. Existing methods for nonlinear transformations can be resource-intensive \cite{rattew2023nonlineartransformationsquantumamplitudes}. Instead, we propose low-degree polynomial approximations of arbitrary nonlinearities, which can be implemented using relatively shallow quantum circuits known as Quantum Nonlinear Processing Units (QNPUs), as developed in \cite{lubasch_zoo}, and are more suitable for near-term NISQ devices.

Accordingly, any nonlinearity $\mathcal{G}(u')$ in the functional \eqref{eq:mardsen_1d} is approximated as
\begin{equation}
    \mathcal{G}(u') \approx \sum_{i=0}^{N_{t_n}} c_i\, (u')^i,
\end{equation}
where $N_{t_n}$ denotes the truncation order,  yielding an approximated functional denoted by $\tilde{\mathcal{E}}$. Let $\bbu$ denote the discrete vector representing the field $u$ and the functional $\tilde{\mathcal{E}}[u]$ be written as a function $\tilde{\mathcal{E}}(\bbu)$ .  Accordingly, we define the VQA cost function as follows:
\begin{equation}
\label{eq:cost_cum_energy}
    \mathcal{C}(\lambda_0, \bblam) := \tilde{\mathcal{E}}(\bbu(\lambda_0, \bblam)).
\end{equation}
Each term in $\tilde{\mathcal{E}}$ is further decomposed into components suitable for quantum evaluation. The corresponding circuits, with the control parameters as inputs, are executed independently, and their expectation values are combined classically to estimate the cost function.

 The second phase is concerned with a classical optimization process to iteratively update the control parameters using a gradient-based optimizer, aimed at searching for a critical point of the approximate potential energy (via cost function). The gradients $({\p \mathcal{C}}/{\p \lambda_i} )$ computed via finite differences use the same circuits as that of the cost function. A cost function value based stopping criteria $\mathcal{F}_{\text{opt}}$ terminates the optimization. The optimal parameters ($\lambda_0^*, \bblam^*$) identify the discrete  profile 
 \begin{equation}
        \bbu^* = \lambda_0^*\, \widehat{V}(\bblam^*) \ket{0},
 \end{equation}
 which corresponds to a critical point of 
$\mathcal{\tilde{E}}$ and is therefore taken as the solution. 

\section{Application to a special case of Neo-Hookean Material Model}
\label{sec:NeoHookean_developement}
   The strain energy density function for a Neo-Hookean material model is given by
\begin{equation}
W(\bbF)
= \frac{\mu}{2}\left( \operatorname{tr}\!\left(\bbF^\top \bbF\right) - \text{ndim} \right)
- \mu \ln J
+ \frac{\Lambda}{2}\big(\ln J\big)^2,
\end{equation}
where $\mu$ and $\Lambda$ are Lame's parameters and ndim represents the number of dimensions. or the present study, we consider a special case of the compressible Neo-Hookean model by setting $\Lambda=0$, which eliminates the volumetric resistance term, and focusing on a unidirectional and one-dimensional representation. This simplifying choice does not limit generality of the proposed approach and an extension of the present work \cite{kouskiya2026hyperelasticity} considers additional forms of nonlinearity. Upon invoking \eqref{eq:def_grad_1}, the strain energy density function reduces to a function of the principal stretch $\lambda$ only:  
\begin{equation}
  W(\lambda) 
= \frac{\mu}{2} \big( \lambda^2 -1 - 2 \ln \lambda \big) 
\label{eq:1d_strain}
\end{equation}
The equilibrium equation \eqref{eq:1d_equilibrium} reduces to:
\begin{equation}
    \diff{}{X} \left( \mu\left(1 + u' - \big(1 + u'\big)^{-1}\right)\right) + B = 0 \quad \text{in } \Omega_0
    \label{eq:strong_form_neo}
    \end{equation}
where
\begin{equation}
     P = \mu \left( \lambda - \frac{1}{\lambda} \right)\mbox{ and} \quad \lambda = 1 + u',
\end{equation}
were used. We employ the following boundary conditions:
\begin{subequations}
\begin{align}
    u(0) &= \bar{u}, \label{eq:bc1} \\
    P(L) &= \bar{P}, \label{eq:bc2}
\end{align}
\label{eq:bcs}
\end{subequations}
where the boundary condition $P(L) = \bar{P}$ implies
\begin{equation}
1 + u'(L) - \left(1 + u'(L)\right)^{-1} = \frac{\bar{P}}{\mu}.
\end{equation}
The corresponding potential energy functional $\mathcal{E}$ can be written explicitly in $u$:
\begin{equation}
\label{eq:energy_func}
    \mathcal{E}[u] = \int_{\Omega_0} \left( 
    \mu \left( u' + \tfrac{1}{2}\left(u'\right)^2 
    - \ln\!\big(1 + u'\big) \right) - B\,u 
    \right)\, \mathrm{d}X 
     -  \bar{P}\,u(L),
\end{equation}
where $u'(X) > -1 \,\forall\, X\in\Omega_0$.
 In this case, it can be shown that for admissible displacements satisfying $u'(X)>-1$, the second variation of the energy functional is strictly positive. Consequently, the functional is strictly convex on the admissible set, implying the existence of a unique minimizer. This minimizer coincides with the unique weak solution of the governing equation and, under sufficient regularity, satisfies the strong form \eqref{eq:strong_form_neo}.

\subsection{QC Compatible Approximation of the Energy} \label{sec:energy_approximations}
\begin{figure}[htbp]
    \centering
    \includegraphics[width=0.6\linewidth]{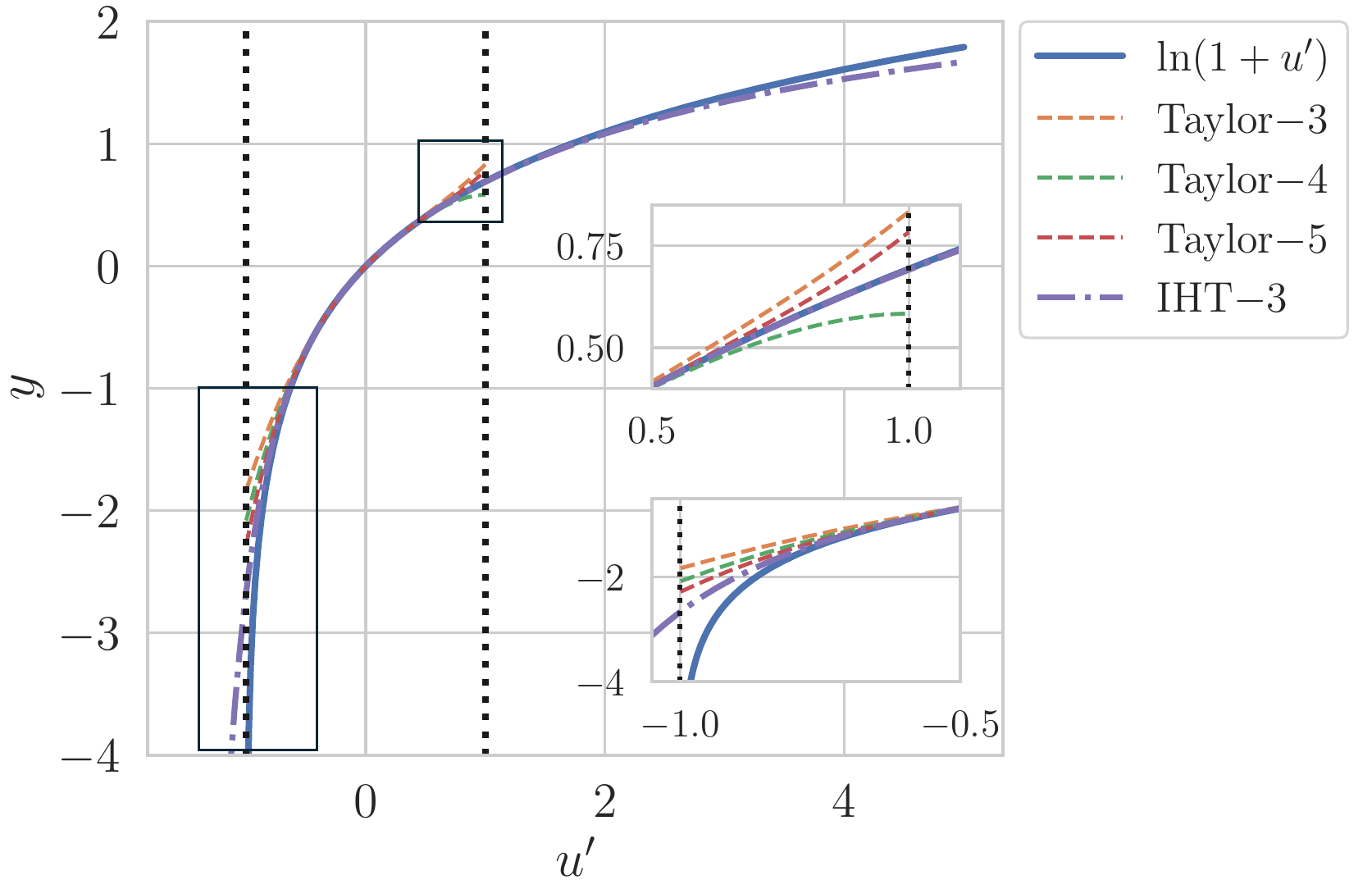}
    \caption{Approximations to the logarithmic nonlinearity. The Taylor series expansion is valid only within $-1<u'<1$. $y$ represents different approximations.}
    \label{fig:ln_uprime_approximations}
\end{figure}

Evaluating \eqref{eq:energy_func} within a quantum framework would require us to deal with the logarithmic nonlinear term. As an illustrative example, the logarithmic term appearing
in~\eqref{eq:energy_func} may be approximated using a Taylor series expansion,
\begin{subequations}
\begin{gather}
    \ln\!\left( 1 + u' \right)
    \approx
    u'
    - \frac{1}{2} \left(u'\right)^2
    + \frac{1}{3} \left(u'\right)^3
    - \cdots
     + \frac{(-1)^{N_{t_n}+1}}{N_{t_n}} \left(u'\right)^{N_{t_n}};\label{eq:log_approximation} \\
     -1 <u'<1 \label{eq:diap_grad_constraint}
     \end{gather}
\end{subequations}
Substituting this approximation into the original energy functional
\eqref{eq:energy_func} yields the truncated polynomial energy
\begin{equation}
\label{eq:energy_upto_cub}
    \tilde{\mathcal{E}}[u]
    =
    \mathcal{E}_B[u]
    + \mathcal{E}_2[u]
    + \sum_{i\geq 3}^{N_{t_n}} \mathcal{E}_i[u] -  \bar{P}\,u(L),
\end{equation}
where the individual contributions are given by
\begin{subequations}
\label{eq:terms_in_energy}
\begin{align}
    \mathcal{E}_B[u]
    &=
    - \int_{\Omega_0} B\, u \, \dd X, \\[4pt]
    \mathcal{E}_2[u]
    &=
    \int_{\Omega_0} \mu \left(u'\right)^2 \, \dd X, \\[4pt]
    \mathcal{E}_i[u]
    &=
    \frac{(-1)^i}{i}
    \int_{\Omega_0} \mu \left(u'\right)^i \, \dd X.
\end{align}
\end{subequations}
The approximation \eqref{eq:log_approximation} introduces an error in the 
constitutive modeling and, consequently, in the computed displacement field. 
For a fixed $|u'|<1$, the strain-energy-density truncation error decays as 
$\mathcal{O}\!\left(|u'|^{N_{t_n}+1}/(N_{t_n}+1)\right)$, while the 
corresponding stress and displacement errors converge geometrically as 
$\mathcal{O}\!\left(|u'|^{N_{t_n}}\right)$; the associated bounds and their 
propagation to the solution are developed in Appendix 
\ref{app:constitutive_error}. Since $u'(X)$ is a spatially varying field, this approximation is used only when $\sup_{X\in\Omega_0}|u'(X)|<1$ is expected to hold over the domain, per \eqref{eq:diap_grad_constraint}. The corresponding loading condition has been developed in Appendix \ref{app:constitutive_error}.
For all other cases, a more suitable representation of the
logarithmic term may be obtained using the inverse hyperbolic tangent (IHT)
expansion~\cite{AbramowitzStegun1964},
\begin{equation}
\label{eq:IHT_approximation}
    \ln\!\left(1 + u'\right)
    \approx
    2\left(
        \frac{u'}{u'+2}
        + \frac{1}{3}\left( \frac{u'}{u'+2} \right)^3
        + \cdots
        + \frac{1}{2\,N_{H_n}-1}\left( \frac{u'}{u'+2} \right)^{2N_{H_n}-1}
    \right),
\end{equation}
where $N_{H_n}$ denotes the number of terms considered in the expansion. This approximation is valid for all $u'>-1$, including the transition point $u'=1$, where the IHT argument $u'/(u'+2)=1/3$ lies well within its convergence radius. For a fixed physically admissible $u'>-1$, the strain-energy-density 
truncation error decays as
$\mathcal{O}\!\left(\left|u'/(u'+2)\right|^{2N_{H_n}+1}/(2N_{H_n}+1)\right)$,
while the corresponding stress and displacement errors converge geometrically 
as $\mathcal{O}\!\left(\left|u'/(u'+2)\right|^{2N_{H_n}}\right)$; the 
associated bounds and their propagation to the solution are developed in 
Appendix \ref{app:constitutive_error}. Although more flexible than the Taylor series, the IHT expansion is costlier to implement within the current framework, requiring the auxiliary variable and penalty formulation introduced below. Consequently, the Taylor expansion is preferred whenever small enough loadings keep $u'$ within its range of validity, with the IHT expansion reserved for larger deformations. 
To incorporate this expansion while maintaining a polynomial energy
representation, we introduce an auxiliary variable
$y$ and define the augmented energy functional
\begin{equation}
\label{eq:penaly_functional}
    \tilde{\mathcal{E}}[u,y]
    =
    \int_{\Omega_0}
    \biggl[
        \mu \left(u'
        + \frac{1}{2}\left(u'\right)^2
        - 2\left( y + \frac{1}{3} y^3 \right) \right)
        - B\,u
        + \frac{\mathcal{P}}{2}\left( u' y + 2y - u' \right)^2
    \biggr] \, \dd X -  \bar{P}\,u(L),
\end{equation}
where $\mathcal{P} \gg 1$ is the penalty parameter enforcing the constraint
\begin{equation}
\label{eq:y_f(u)}
y \approx {u'}/{(u'+2)}
\end{equation}
during minimization (the expansion has been truncated after two terms). In addition to the IHT truncation error, the finite-penalty enforcement introduces a separate $\mathcal{O}(\mathcal{P}^{-1})$ error in the strain energy density, stress, and resulting displacement field, which vanishes as $\mathcal{P}\rightarrow\infty$ (see Appendix \ref{app:constitutive_error}). The above functional is amenable to implementation within a quantum framework which is explained in the following subsections. The resulting variational problem is to determine
\begin{equation}
    (u^*, y^*)
    =
    \arg\min_{u,y} \, \tilde{\mathcal{E}}[u,y],
    \qquad
    \text{subject to } u(0) = \bar{u} .
\end{equation}
Fig.~\ref{fig:ln_uprime_approximations} compares the accuracy of the two types of expansions.

\subsection{Finite Element Discretization}
\label{sec:discrete_integrals}

We employ a Ritz-type approximation in which the solution field is expressed as a finite expansion in terms of trial (basis) functions \cite{Rektorys1977}. In the present work, these trial functions coincide with standard finite-element shape functions. This section presents the discrete integral structure for various terms in \eqref{eq:energy_func}, arising from the polynomial order of basis functions. This structure is then exploited to design the quantum circuits.

Let the domain $\Omega_0$ be discretized into finite elements with $N_q + 1$
nodal points such that $N_q = 2^n$ where $n$ represents the number of main qubits (one node is reserved as Dirichlet boundary condition). The number of elements depends on the polynomial order of the shape functions employed; for instance, $N_q$ linear elements or $N_q/2$ quadratic elements. 

The finite element approximation of an arbitrary field $f$ is given by
\begin{equation}
    \tilde{f}(X) := \sum_{i=0}^{N_q} f_i\, N_i(X) = \bbf^\top \boldsymbol{N},
\end{equation}
where $N_i(X)$ denotes the Lagrange shape function associated with node $i$, and $f_i$ is the corresponding nodal value.

Within an element $e$, the FE approximation can be written in local form as
\begin{equation}
    \tilde{f}(X(\xi)) = \sum_{i=0}^{N_{en}-1} f_{i_e}\, \widehat{N}_i(\xi),
\end{equation}
where $i$ indexes the $N_{en}$ local nodes of element $e$ and $i_e$ represents the global node number associated with the local node $i$. In this work, 
\begin{equation}
    i_e = \begin{cases}
    i + e \quad &\mbox{ for } 1^{\mathrm{st}}\mbox{-order shape functions};
    \\ 
    i + 2e \quad &\mbox{ for } 2^{\mathrm{nd}}\mbox{-order shape functions}.
    \end{cases}
\end{equation}
The derivative of $\tilde{f}$ with respect to the physical coordinate $X$,
restricted to element $e$, is
\begin{equation}
    \tilde{f}'(X) = \sum_{i=0}^{N_{en}-1} f_{i_e}\, N_{i_e}'(X),
\end{equation}
where the derivatives of the shape functions follow from the chain rule:
\begin{equation}
    N_{i_e}'(X) = \widehat{N}_i'(\xi(X))\,\big(J_e(\xi(X))\big)^{-1},  
    \qquad J_e(\xi) = \diff{X}{\xi}.
\end{equation}
Here $J_e(\xi)$ is the Jacobian of the mapping from the reference element to
physical element $e$, and $\xi(X)$ denotes the inverse mapping within element $e$.
  
The integral of $\tilde{f}$ over the element $\Omega_e$ is approximated via Gaussian quadrature as
\begin{equation}
    \int_{\Omega_{e}} \tilde{f}(X) \, \mathrm{d}X 
    \;\approx\; \sum_{g=0}^{N_{gp}-1} w^{(g)} \, J_e^{(g)} \, \tilde{f}\big(X^{(g)}\big),
\end{equation}
where $N_{gp}$ is the number of Gauss points in the element, $w^{(g)}$ is the quadrature weight at Gauss point $g$, $J_e^{(g)}$ is the Jacobian evaluated at $\xi^{(g)}$ for element $e$, and $X^{(g)} = X_e(\xi^{(g)})$ is the corresponding physical coordinate in $\Omega_e$.

\subsubsection{First-order FE shape functions}
For first-order FE shape functions and two-point Gauss quadrature, let
\begin{equation}
\label{eq:SF1_interpolation}
    \alpha_i^{(I,g)} =\widehat{N}_i\big(\xi^{(g)}\big) \quad \mbox{ and } \quad
    \alpha_i^{(II,g)} = \widehat{N}_i'(\xi^{(g)}).
\end{equation}
Accordingly, $\tilde{f}\big(X^{(g)}\big)$ reduces to:
\begin{equation}
\label{eq:gauss_approx}
\begin{gathered}
    \tilde{f}\big(X^{(0)}\big) = f_{0_{e}} \alpha_0^{(I,0)} + f_{1_{e}} \alpha_1^{(I,0)} = \alpha \, f_{0_{e}} + (1-\alpha) \,f_{1_{e}}; \\
    \tilde{f}\big(X^{(1)}\big) = f_{0_{e}} \alpha_0^{(I,1)} + f_{1_{e}} \alpha_1^{(I,1)}  = (1-\alpha) \, f_{0_{e}} + \alpha\, f_{1_{e}}; \\
     \alpha = \frac{\sqrt{3} + 1}{2\sqrt{3}}. 
\end{gathered}
\end{equation}
Similarly, the derivative of $\tilde{f}$ within any element $e$ and at Gauss point $g$ can be approximated as:
\begin{equation} 
    \tilde{f}'(X^{(g)}) = \sum_{i=0}^{1} f_{i_e}\, N_{i_e}'(X^{(g)}) = \sum_{i=0}^{1} \frac{2\, f_{i_e}}{h_e}\,\alpha_i^{(II,g)}  = \frac{f_{1_e} - f_{0_e}}{h_e},
    \label{eq:deriv_ele_approx}
\end{equation}
where $h_e$ denotes the length of  element $e$.
Using \eqref{eq:gauss_approx} and \eqref{eq:deriv_ele_approx}, each of the terms in \eqref{eq:terms_in_energy} can be approximated via the following element-wise quadrature evaluations:
    \begin{subequations}
\begin{gather}
    \label{eq:quad_energy}\tilde{\mathcal{E}}_2(\boldsymbol{u})=  \mu\,\bigintsss_{\Omega_0}\left(\tilde{u}'\right)^2 \,\dd X = \mu \sum_{e=0}^{N_q - 1} h_e \left(\frac{u_{1_e} - u_{0_e}}{h_e}\right)^2;\\
    \label{eq:cub_energy} \tilde{\mathcal{E}}_i(\boldsymbol{u})=  \mu\,\bigintsss_{\Omega_0}\frac{(-1)^i}i\left(\tilde{u}'\right)^i \, \dd X = \, \mu\left(\frac{(-1)^{i}}i\right) \sum_{e=0}^{N_q - 1} h_e \left(\frac{u_{1_e} - u_{0_e}}{h_e}\right)^i;\\
    \label{eq:body_energy} \tilde{\mathcal{E}}_B(\boldsymbol{u})= -  \int_{\Omega_0}\,B\,\tilde{u} \,\dd X =  - \sum_{e=0}^{N_q - 1}\sum_{g=0}^{1}
\frac{h_e}{2}\,
B\!\left(X_e^{(g)}\right)\,
\left(\sum_{i=0}^{1} u_{i_e}\,\alpha_i^{(I,g)}\right).
\end{gather}
\label{eq:energy_terms}
\end{subequations}

A similar discrete decomposition for the functional
\eqref{eq:penaly_functional}
is presented in Appendix~\ref{app:penalty_formulation}.

\subsubsection{Second-order Shape Functions} \label{sec:second_order_discretizzation}
For simplicity, the central node within an element is placed at the center of the element: $X_{1_e} = (X_{0_e} + X_{2_e})/2.$
As a consequence, the value of the Jacobian over the length of the domain becomes a constant: $J_e(\xi) = {h_e}/{2}$. For second-order FE shape functions and any Gauss point g, we introduce
\begin{equation}
\label{eq:SF2_interpolation}
    \beta_i^{(I,g)} := \widehat{N}_i\big(\xi^{(g)}\big) \quad \mbox{and} \quad \beta_i^{(II,g)} = \widehat{N}_i'(\xi^{(g)}).
\end{equation}
The value of the function $\tilde{f}$ is expressed as
\begin{equation}
\label{eq:beta_I_formulas}
  \tilde{f}\big(X^{(g)}\big) = \sum_{i=0}^{2} f_{i_e}\, \beta_i^{(I,g)}.
\end{equation}
where the values of $\beta_i^{(I,g)}$ are established in a straightforward way. Similarly, the derivative is approximated as:
\begin{equation}
\tilde{f}'\big(X^{(g)}\big) = \sum_{i=0}^2 f_{i_e}\, N_{i_e}'(X^{(g)}) = 2\sum_{i=0}^{2} \frac{f_{i_e}\,\beta_i^{(II,g)}}{h_e}.
\end{equation}
For a two-point Gauss quadrature (for simplicity) and $N_{el} = N_q/2$ elements, we obtain the following:
\begin{subequations}
\begin{gather}
\label{eq:SF2_quad_energy}
\tilde{\mathcal{E}}_2(\boldsymbol{u})
=
\mu
\sum_{e=0}^{N_{el}-1}
\sum_{g=0}^{1}
\left(\frac{2}{h_e}\right)
\left(
\sum_{j=0}^{2} u_{j_e}\,\beta_j^{(II,g)}
\right)^2, \\
\label{eq:SF2_cub_energy}
\tilde{\mathcal{E}}_i(\boldsymbol{u})
=
\mu\left(\frac{(-1)^{i}}{i}\right)
\sum_{e=0}^{N_{el}-1}
\sum_{g=0}^{1}
\left(\frac{2}{h_e}\right)^{i-1}
\left(
\sum_{j=0}^{2} u_{j_e}\,\beta_j^{(II,g)}
\right)^i, \\
\label{eq:SF2_body_energy}
\tilde{\mathcal{E}}_B(\boldsymbol{u})
=
-
\sum_{e=0}^{N_{el}-1}
\sum_{g=0}^{1}
\frac{h_e}{2}
\left(
\sum_{j=0}^{2} u_{j_e}\,\beta_j^{(I,g)}
\right)
B\!\left(X_e^{(g)}\right).
\end{gather}
\label{eq:SF2_energy_terms}
\end{subequations}

\section{Implementation of the Cost Function}
\label{sec:VQA_for_neohookean}

This section focuses on the construction of the approximate discretized energy functional $\tilde{\mathcal{E}}$ using quantum circuits. We begin by introducing the circuits that serve as the fundamental building blocks for evaluating the cost function. We then describe the state-preparation procedure used to encode real-valued vectors as quantum states. Finally, we show how these components are combined to evaluate the discrete integrals using quantum circuits, thereby enabling the computation of the cost function.

\subsection{Algorithmic Primitives}
\label{sec:building_blocks}

All quantum circuits employed in this work are constructed within a common framework built upon the idea of a Quantum Nonlinear Processing Unit (QNPU)~\cite{lubasch_zoo}. The template circuit representing a framework for the calculations, referred to as the QNPU framework, is shown in Fig.~\ref{fig:VQA_outline}. Circuits with different QNPU blocks enable the evaluation of scalar quantities of the form
\begin{equation}
\label{eq:QNPU_operation}
E = \hat{f}^{(1)\,\top}\left( \odot_{j=1}^{r} \, O_j \hat{f}^{(j)} \right),
\end{equation}
where $O_j$ denotes a linear operator acting on the input quantum state $\hat{f}^{(j)}$, and $\odot$ represents the element-wise (Hadamard) product across the vectors $\{O_j \hat{f}^{(j)}\}_{j=1}^r$. The input states $\hat{f}^{(j)}$ are prepared on separate quantum registers (input ports), which are subsequently processed by the QNPU block through entangling operations. Following this processing stage, $E$ is measured using an ancilla qubit, without requiring full state tomography. In what follows, we refer to this ancilla as the \textit{Hadamard ancilla}, to distinguish it from other types of ancilla qubits. In particular,
\begin{equation}
\label{eq:ancilla_QNPU_read}
E = \mathbb{P}(0) - \mathbb{P}(1),
\end{equation}
where $\mathbb{P}(0)$ and $\mathbb{P}(1)$ are the probabilities of measuring $\ket{0}$ and $\ket{1}$ on the Hadamard ancilla, respectively.

\subsubsection{Inner Product via QNPU}
\label{sec:inner_product}
Let $\bbp, \bbq \in \mathbb{R}^{N_q}$ be real vectors parameterized by control parameters $(\lambda_{p_0}, \boldsymbol{\lambda}_p)$ and $(\theta_{q_0}, \boldsymbol{\theta}_q)$, respectively, for a fixed ansatz:
\begin{equation}
\bbp = \|\bbp\| \ket{\hat{p}} = \lambda_{p_0} \, \widehat{V}(\boldsymbol{\lambda}_p) \ket{0}, 
\quad
\bbq = \|\bbq\| \ket{\hat{q}} = \theta_{q_0} \, \widehat{V}(\boldsymbol{\theta}_q) \ket{0}.
\end{equation}
The inner product between $\bbp$ and $\bbq$ is given by
\begin{equation}
\label{eq:inner_prod}
\bbp^{\top} \bbq
=
\|\bbp\| \, \|\bbq\| \,
\braket{\hat{p}}{\hat{q}}
=
\lambda_{p_0} \theta_{q_0}
\left\langle 0 \middle|
\widehat{V}^\dagger(\boldsymbol{\lambda}_p)\,
\widehat{V}(\boldsymbol{\theta}_q)
\middle| 0
\right\rangle.
\end{equation}

\begin{figure}[htbp]
    \centering
    
    \begin{subfigure}[b]{0.45\textwidth}
        \centering
        \includegraphics[width=\textwidth]{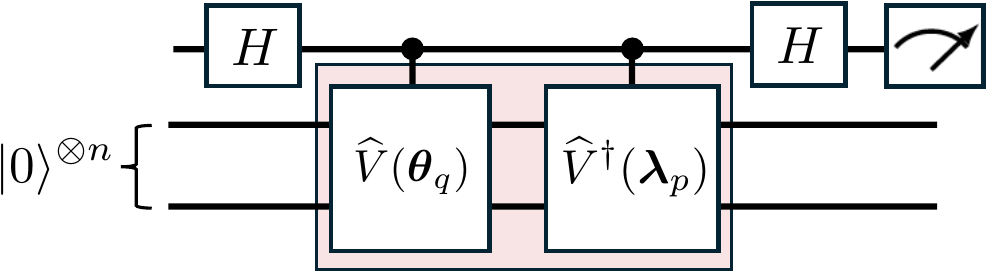}
        \caption{}
        \label{fig:inner_prod}
    \end{subfigure}
    \hfill
    \begin{subfigure}[b]{0.45\textwidth}
        \centering
        \includegraphics[width=\textwidth]{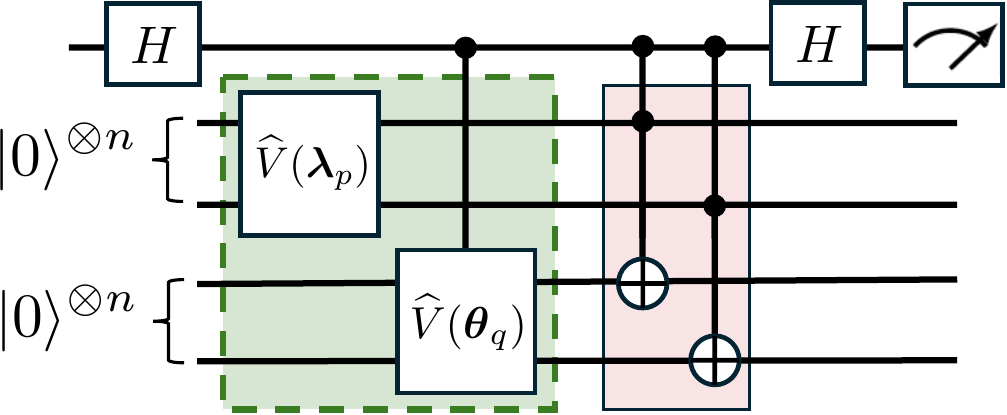}
        \caption{}
        \label{fig:diagonal_circuit}
    \end{subfigure}

    \vspace{0.8cm}

    \begin{subfigure}[b]{0.45\textwidth}
        \centering
        \includegraphics[width=\textwidth]{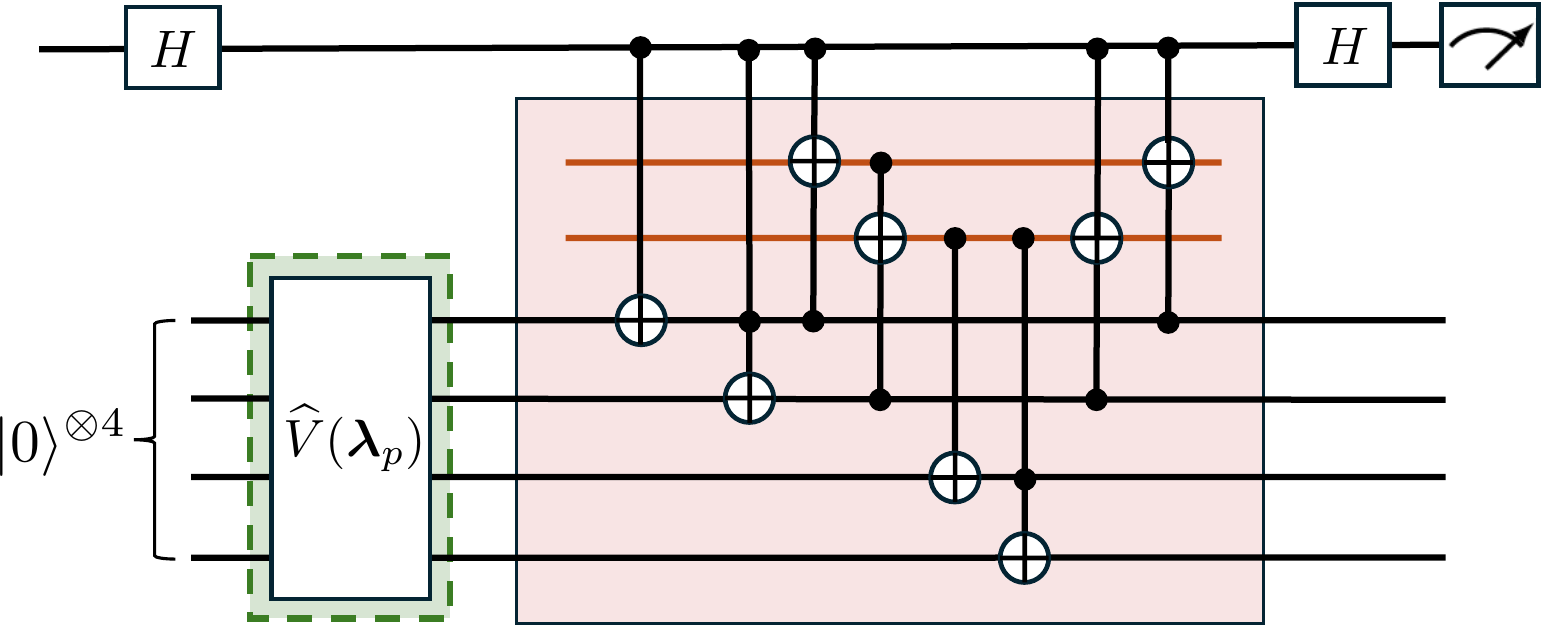}
        \caption{}
        \label{fig:adder_full}
    \end{subfigure}
    \hfill
    \begin{subfigure}[b]{0.45\textwidth}
        \centering
        \includegraphics[width=\textwidth]{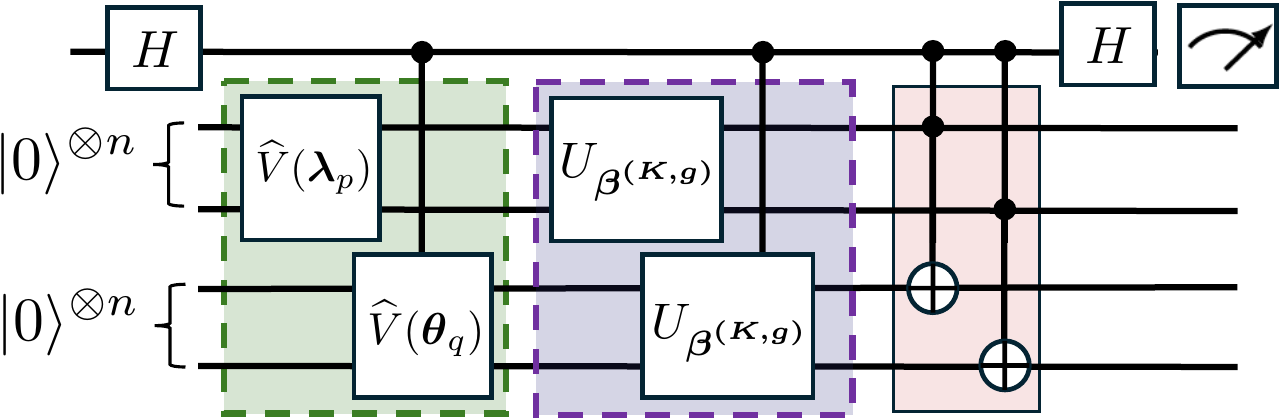}
        \caption{}
        \label{fig:block_encode_qnpu}
    \end{subfigure}

    \caption{
Primitive circuits to evaluate  
(a)~$\braket{\hat{p}}{\hat{q}}$, 
(b)~$\bra{\hat{p}} D_{\hat{q}}\ket{\hat{p}}$, 
(c)~$\bra{\hat{p}}\widehat{A}\ket{\hat{p}}$, and 
(d)~$\|\boldsymbol{\ell}_p\|^2 \, \|\boldsymbol{\ell}_q\|
\, \bra{\hat{\ell}_p} D_{\hat{\ell}_q} \ket{\hat{\ell}_p}$; The green, red, and purple regions denote the input, QNPU, and block-encoding phases, respectively. In (c), the two red qubit registers correspond to the ancillas required by the Adder circuit. In (d), ancillas associated with the block-encoding components are omitted for clarity.
}
    \label{fig:building_blocks_all}
\end{figure}

Comparing \eqref{eq:inner_prod} with \eqref{eq:QNPU_operation} (up to the scaling factor and setting $r=1$), we identify $\ket{\hat{f}^{(1)}} = \ket{0}$, which, being the initial computational basis state, does not require state preparation. The corresponding operator is given by
\begin{equation}
O_1 = \widehat{V}^\dagger(\boldsymbol{\lambda}_p)\,\widehat{V}(\boldsymbol{\theta}_q),
\end{equation}
which is implemented within the QNPU framework, as illustrated in Fig.~\ref{fig:inner_prod}. This circuit requires $n+1$ qubits.

\subsubsection{Diagonal Operators}
\label{sec:diagonal_QNPU}
To evaluate quantities involving element-wise (Hadamard) products of more than two quantum state vectors, we introduce diagonal operators. As an illustrative example, consider the quantity
\begin{equation}
\label{eq:diagonal_measure}
\bbp^\top (\bbq \odot \bbp),
\end{equation}
which can be expressed as
\begin{equation}
\label{eq:diagonal_QNPU_form}
\bbp^\top (\bbq \odot \bbp)
= \|\bbp\|^2 \, \|\bbq\| \sum_{i=0}^{N_q-1} \hat{p}_i^2 \hat{q}_i
= \lambda_{p_0}^2 \theta_{q_0} \, \bra{\hat{p}} D_{\hat{q}} \ket{\hat{p}},
\end{equation}
where $D_{\hat{q}}$ is a diagonal operator defined by
\begin{equation}
(D_{\hat{q}})_{ii} = \hat{q}_i.
\end{equation}
For simplicity, we will often use
\begin{equation}
(D_{\bbq})_{ii} = \|\bbq\| \,\hat{q}_i.
\end{equation}
The action of $D_{\hat{q}}$ on the state $\ket{\hat{p}}$ is given by
\begin{equation}
D_{\hat{q}} \ket{\hat{p}} = \sum_{i=0}^{N_q-1} \hat{q}_i \hat{p}_i \ket{i},
\end{equation}
which corresponds to element-wise (Hadamard) scaling of the amplitudes of $\ket{\hat{p}}$ by those of $\ket{\hat{q}}$.

The circuit diagram to evaluate  \eqref{eq:diagonal_QNPU_form}, up to the scaling factor is shown in Fig.~\ref{fig:diagonal_circuit}. Comparing \eqref{eq:diagonal_QNPU_form} with \eqref{eq:QNPU_operation}, we identify $\ket{\hat{f}^{(1)}} = \ket{\hat{p}}$, prepared independently of the Hadamard ancilla in input port~1, while $\ket{\hat{f}^{(2)}}$ is prepared in input port~2 under Hadamard ancilla control. These inputs are combined within the QNPU through the action of the diagonal operator. The expectation value obtained from measuring the Hadamard ancilla (cf.~\eqref{eq:ancilla_QNPU_read}) yields the quantity in \eqref{eq:diagonal_measure}, up to a normalization factor. Additional input ports may be appended and entangled with input port 1 in the same manner as input port 2, yielding expressions of the form
\begin{equation}
    \bbp^\top \left(\bbq_1 \odot \bbq_2 \odot \cdots \odot \bbq_{N_t} \odot \bbp \right),
\end{equation}
where \(\bbq_i \in \mathbb{R}^{N_q}, \, i=\{1,2,\cdots N_t\}\),  represent additional input vectors. The corresponding circuit for evaluating the above expression employs \(N_t+1\) input ports and a total of \(n(N_t+1)+1\) qubits.

\subsubsection{Adder Circuit}
The adder circuit $\widehat{A}$ acts on a state $\ket{\hat{p}}$ as a cyclic shift operator, producing
$
\widehat{A}\ket{\hat{p}} = \sum_{k=0}^{N_q - 1} \hat{p}_{k+1} \ket{k},
$
where the index shift is understood modulo $N_q$ \cite{NielsenChuang2010,lubasch_zoo}. This operation can be implemented using quantum circuits that typically require $n-2$ ancillas. Alternatively, an ancilla-free implementation based on the Quantum Fourier Transform (QFT) can be used, albeit at the cost of increased circuit depth \cite{10.1145/502090.502097}.

When embedded within the QNPU framework (see Fig.~\ref{fig:adder_full}), a total of $2n-1$ qubits are required. This circuit enables the evaluation of inner products involving shifted vectors. In particular,
\begin{equation}
\bbp^\top (\widehat{A}\bbp)
= \|\bbp\|^2 \,\bra{\hat{p}} \widehat{A} \ket{\hat{p}}
= \lambda_{p_0}^2 \bra{\hat{p}} \widehat{A} \ket{\hat{p}}.
\end{equation}

In circuit constructions combining multiple primitive operations, several adder circuits may be required. Although the ancillas used in each adder operation evolve during computation, they are ultimately uncomputed and restored to the $\ket{0}$ state. Consequently, the same ancilla register can be reused across successive controlled or uncontrolled adder circuits, thereby reducing the overall ancilla overhead.

\subsubsection{Block Encoding within QNPU}
\label{sec:block_encode}
Quantum circuits implement only unitary transformations, whereas many operators of interest are generally non-unitary. To realize such operators within a quantum circuit framework, we use \emph{block encoding}, where a non-unitary matrix is embedded into a larger unitary operator acting on an extended Hilbert space.

For an operator $\bbA \in \mathbb{C}^{2^s \times 2^s}$, an exact block encoding consists of a unitary $\widehat{U}_{\bbA}$ acting on the system together with ancillary qubits such that
\begin{equation}
\left( \langle 0^{\otimes a}| \otimes I \right)
\widehat{U}_{\bbA}
\left( |0^{\otimes a}\rangle \otimes I \right)
=
\frac{\bbA}{\alpha},
\end{equation}
where \(\alpha > 0\) is a normalization factor such that the scaled operator \(\bbA/\alpha\) can be embedded within a unitary operator and \(a\) is the number of ancilla qubits.

Equivalently, when the ancilla register is initialized in the state \(\ket{0}^{\otimes a}\),
\begin{equation}
\widehat{U}_{\bbA} \left( \ket{0}^{\otimes a} \otimes \ket{\psi} \right)
=
\frac{1}{\alpha}\ket{0}^{\otimes a}\otimes (\bbA \ket{\psi})
+ \ket{\Phi^\perp},
\end{equation}
where \(\ket{\Phi^\perp}\) has no support on ancilla state \(\ket{0}^{\otimes a}\). Thus, postselecting the ancilla register onto the state \(\ket{0}^{\otimes a}\) produces a quantum state proportional to \(\bbA\ket{\psi}\), where the corresponding proportionality factor \(\|\bbA\ket{\psi}\|\) is determined by the success probability of the postselection measurement. 

For notational simplicity, ancilla registers and postselection operations will be omitted whenever implicit. Accordingly, we adopt the notation
\begin{equation}
\label{eq:Block_encode_notation}
U_{\bbA}\ket{\psi}
\equiv
\|\bbA\ket{\psi}\|
\left(
\frac{\bbA\ket{\psi}}{\|\bbA\ket{\psi}\|}
\right)
\end{equation}
to denote the effective non-unitary action of \(\bbA\) on \(\ket{\psi}\) realized through the block encoding \(\widehat{U}_{\bbA}\), where the bracketed term corresponds to the normalized quantum state obtained on the system registers following postselection.

In this work, we follow the construction in \cite{osti_2466156} to implement block encodings of non-unitary sparse circulant matrices. For non-circulant operators, which are expected to arise in higher-dimensional finite-element discretizations, hierarchical Pauli decomposition techniques~\cite{philo2026phase} may offer an alternative. In particular, we use \eqref{eq:SF1_interpolation} and \eqref{eq:SF2_interpolation} to define the matrices  $\boldsymbol{\alpha^{(K,g)}}, \boldsymbol{\beta^{(K,g)}}  \in \mathbb{R}^{N_q \times N_q}, K = \{I,II\}$, such that
\begin{equation} \label{eq_circulant_matrix_I}
\left(\boldsymbol{\alpha^{(K,g)}}\right)_{ij}
=
\begin{cases}
\alpha^{(K,g)}_0, & j \equiv i \pmod{N_q}, \\[4pt]
\alpha^{(K,g)}_1, & j \equiv i+1 \pmod{N_q}, \\[4pt]
0, & \text{otherwise};
\end{cases}
\quad i,j = 1,\dots,N_q,
\end{equation} 
and 
\begin{equation} \label{eq_circulant_matrix_II}
\left(\boldsymbol{\beta^{(K,g)}}\right)_{ij}
=
\begin{cases}
\beta^{(K,g)}_0, & j \equiv i \pmod{N_q}, \\[4pt]
\beta^{(K,g)}_1, & j \equiv i+1 \pmod{N_q}, \\[4pt]
\beta^{(K,g)}_2, & j \equiv i+2 \pmod{N_q}, \\[4pt]
0, & \text{otherwise};
\end{cases}
\quad i,j = 1,\dots,N_q.
\end{equation}
The block encodings of \(\boldsymbol{\alpha^{(K,g)}}\) and \(\boldsymbol{\beta^{(K,g)}}\) require \(2n\) and \(2n+1\) qubits, respectively, where \(n\) denotes the number of qubits in the main register. Each block-encoding circuit incorporates one or more adder circuits, allowing the associated \(n-2\) adder ancilla qubits to be shared and reused within and across block encodings. In addition, the block encodings of \(\boldsymbol{\alpha^{(K,g)}}\) and \(\boldsymbol{\beta^{(K,g)}}\) require \(2\) and \(3\) dedicated ancilla qubits, respectively, for postselection measurements. These ancillas cannot be reused once measured.

Embedding these block-encoded operators within the QNPU framework enables the evaluation of more complex expressions. For example, let $\bbm\in\mathbb{R}^{N_q}$ such that
\begin{equation}
\label{eq:Block_encode_action_q}
\boldsymbol{\beta^{(K,g)}}\bbm = \|\bbm\|\, U_{\boldsymbol{\beta^{(K,g)}}} \ket{\hat{m}} = \|\bbm\| \boldsymbol{\ell}_m^{(K,g)} = \|\bbm\| \left\|\boldsymbol{\ell}_m^{(K,g)}\right\| \,\left|\hat{\ell}_m^{(K,g)}\right\rangle.
\end{equation}
Then, using \eqref{eq:diagonal_QNPU_form} and \eqref{eq:Block_encode_notation},
\begin{equation}
\left(\boldsymbol{\beta^{(K,g)}}\bbp\right)^\top \left((\boldsymbol{\beta^{(K,g)}}\bbq) \odot (\boldsymbol{\beta^{(K,g)}}\bbp)\right)
= \|\bbp\|^2 \, \|\bbq\|
\,
 \left\langle \hat{p}  \middle| U_{\boldsymbol{\beta^{(K,g)}}}^\dagger  D_{\bbll_q^{(K,g)}} U_{\boldsymbol{\beta^{(K,g)}}} \middle|  \hat{p} \right\rangle,
\label{eq:cubic_block_encode_formula}
\end{equation}
where $U_{\boldsymbol{\beta^{(K,g)}}}^\dagger$ denotes the adjoint block encoding, obtained as the circuit inverse of $U_{\boldsymbol{\beta^{(K,g)}}}$ and thus block-encoding $\left(\boldsymbol{\beta^{(K,g)}}\right)^\top$ rather than requiring any symmetry assumption on the circulant matrices, and the bracketed quantity on the right-hand side can be evaluated using the circuit in Fig.~\ref{fig:block_encode_qnpu}. The corresponding implementation requires a total of \(3n+5\) qubits. Of these, $2n$ qubits correspond to the input registers, $n-2$ ancillas are used for adder operations, and one ancilla serves as the Hadamard ancilla. The remaining six ancillas (three per block encoding) are associated with the block encoding circuits and require postselection measurement.

\subsection{State preparation} \label{sec:state_prep}
The proposed algorithm employs a number of predefined auxiliary vectors, such as body-force fields or element-length information. This data, represented using non-normalized vectors in $\mathbb{R}^{N_q}$, must be incorporated into the quantum framework so that the computed solution reflects the prescribed parameters. 
The quantum state $\ket{\hat{f}}$ of vector $\bbf$ is then encoded into the relevant quantum circuit through a \textit{state preparation} stage. In this work, we employ the same ansatz $\widehat{V}$ circuit used to represent the input state to also prepare the state $\ket{\hat{f}}$, where the required ansatz parameters are determined through a classical optimization procedure. Specifically, we solve
\begin{equation}
\label{eq:state_prep}
    (\lambda_{f_0}, \boldsymbol{\lambda}_f)
    = \arg\min_{\lambda_0,\, \boldsymbol{\lambda}}
    \; \mathcal{C}_f (\lambda_0, \boldsymbol{\lambda}), \quad\mbox{where }\quad \mathcal{C}_f(\lambda_0, \boldsymbol{\lambda})
    := \left\| \bbf
    - \lambda_0 \, \widehat{V}(\bblam)\ket{0} \right\|
\end{equation}
The tolerance for the optimization in the current work (i.e., examples below) is set to $\mathcal{C}_f\leq10^{-5}$. 

\subsection{Evaluating discrete integrals} \label{sec:discrete_integral_circuits}
The discrete integrals in Section~\ref{sec:discrete_integrals} are computed as summations over all elements, where each element involves specific interpolant-based manipulations (e.g., evaluating fields at Gauss points via interpolation of elemental Degrees of Freedom, DoFs) followed by possible nonlinear arithmetic (e.g. raising this result by a polynomial power). While these operations are performed locally in the classical finite element framework, quantum circuits act globally on the field through unitary transformations. Consequently, the local arithmetic structure must be extended to align with the global and unitary nature of the QC framework, ensuring that the classical computations can be performed in the quantum setting. 
To achieve this, we present two approaches: (1)~a direct-expansion approach, and (2)~a block-encoding based approach.

In the examples considered in this work, the mesh contains one more node than the Hilbert-space dimension $ N_q = 2^n $ associated with an $ n $-qubit register. 
This extra node enforces the Dirichlet boundary condition within the classical part of the computation, while the remaining degrees of freedom are encoded in the quantum state vector. For a left Dirichlet boundary condition, the global DoF vector  $\bbu$ is partitioned as
\begin{equation} \label{eq_vector_partition}
\bbu = (\bar{u}, \bbv), \quad 
\bbv = (v_0, v_1, \ldots, v_{N_q-1}) = (u_1, u_2, \ldots, u_{N_q}),   
\end{equation}
and $u_0=\bar{u}$. The vector of internal displacements $\bbv$ is parameterized using:
\begin{equation}
\label{eq:quantum_rep_of_u}
\bbv = \lambda_0 \widehat{V} (\bblam) \ket{0}.
\end{equation}
Analogously, the quantum-encoded vector for a right Dirichlet boundary specification becomes
\begin{equation}
\bbv = (u_0, u_1, \ldots, u_{N_q-1}).
\end{equation}
Application of Dirichlet boundary conditions at both ends follows in a straightforward fashion:
\begin{equation}
\bbu = (\bar{u}_0,\, \bbv, \,\bar{u}_{N_q+1}), \quad 
\bbv = (u_1, u_2, \ldots, u_{N_q}),
\end{equation}
The analyses below focus on left Dirichlet condition and a traction boundary condition on the right; however, the other two configurations discussed above can be implemented based on the same ideas presented below.

\subsubsection{Direct-expansion based method}
\label{sec:Direct_expansion}
Considering \eqref{eq:energy_upto_cub}, with the element-wise quadrature evaluations given in \eqref{eq:energy_terms}, we now present the quantum realization of each term:

The \emph{quadratic term} \eqref{eq:quad_energy} is rewritten as
\begin{equation}
E_2 = \mu \sum_{e=0}^{N_q-1} h_e \left(\frac{u_{1_e} - u_{0_e}}{h_e}\right)^2 = \mu\left(\frac{\bar{u}^2}{h_0} + T_1 + T_2 -  T_3\right)
\end{equation}
where
\begin{equation}
T_1 = -\,\frac{2\bar{u}}{h_0}\,v_0, \hspace{0.3cm} T_2 = \sum_{i=0}^{N_q-2} v_i^2\!\left(\frac{1}{h_i} + \frac{1}{h_{i+1}}\right) + \frac{v_{N_q-1}^2}{h_{N_q-1}}, \hspace{0.3cm}
T_3 = \sum_{i=0}^{N_q-2} \frac{2\,v_i\,v_{i+1}}{h_{i+1}} .
\end{equation}

Defining
\begin{equation}
\boldsymbol{m}^{(1)} := \|\boldsymbol{m}^{(1)}\|\,\ket{\hat{m}^{(1)}} = [1,0,0,\ldots,0],
\label{eq:extract_vals}
\end{equation}
and since $\bar{u}$ is the prescribed Dirichlet boundary value at the left end of the domain, the term $T_1$ is evaluated as
\begin{equation}
T_1
= -\,2\,\frac{\bar{u}\,\lambda_0\,\|\boldsymbol{m}^{(1)}\|}{h_0}
\left\langle \hat{v}\middle| \hat{m}^{(1)} \right\rangle,
\end{equation}
where $\|\boldsymbol{m}^{(1)}\|=1$, and the inner-product circuit described in Section~\ref{sec:inner_product} is utilized.

Similarly, $T_2$ can be written as
\begin{equation}
T_2
= \lambda_0^2\,\|\boldsymbol{m}^{(2)}\|
\left\langle \hat{v}\middle| D_{\hat{m}^{(2)}} \middle| \hat{v} \right\rangle,
\end{equation}
with
\begin{equation}
m^{(2)}_i =
\begin{cases}
\dfrac{1}{h_i} + \dfrac{1}{h_{i+1}}, & i = 0,\ldots,N_q-2, \\[6pt]
\dfrac{1}{h_{N_q-1}}, & i = N_q-1 .
\end{cases}
\end{equation}
The circuit described in Section~\ref{sec:diagonal_QNPU} is used to evaluate this expression.

Finally, the cross term $T_3$ can be written explicitly as
\begin{align}
T_3
&= \frac{2\,v_0\,v_{1}}{h_{1}} + \frac{2\,v_1\,v_{2}}{h_{2}} + \cdots  + \frac{2\,v_{N_q - 2}\,v_{N_q - 1}}{h_{N_q - 1}} + 0 \cdot \,v_{N_q - 1}\,v_{0} .
\end{align}
Defining
\begin{equation}
m^{(3)}_i =
\begin{cases}
\dfrac{1}{h_{i+1}}, & i = 0,\ldots,N_q-2, \\[4pt]
0, & i = N_q-1 ,
\end{cases}
\end{equation}
we obtain
\begin{align}
T_3
&= 2\, \bbv^\top \left(\boldsymbol{m}^{(3)} \odot  \widehat{A} \bbv\right) = 2\,\lambda_0^2\,\|\boldsymbol{m}^{(3)}\|
\left\langle \hat{v}\middle| D_{\hat{m}^{(3)}}\,\widehat{A} \middle| \hat{v} \right\rangle,
\end{align}
where a diagonal operator followed by an adder circuit is implemented within the QNPU block; the vanishing mask entry $m^{(3)}_{N_q-1}=0$ suppresses the periodic wrap-around term that the cyclic action of $\widehat{A}$ would otherwise generate, so that the masked coupling reduces to the intended nonperiodic nearest-neighbor interaction on $\Omega_0$. The control parameters corresponding to $\bbm^{(\cdot)}$ are obtained using the state-preparation method described in Section~\ref{sec:state_prep}.

The \emph{cubic term} \eqref{eq:cub_energy} is expressed as
\begin{equation}
E_3 = -\frac{\mu}{3} \sum_{e=0}^{N_q-1} h_e
\left(\frac{u_{1_e}-u_{0_e}}{h_e}\right)^3 = -\frac{\mu}{3}
\left(
-\frac{\bar{u}^3}{h_0^2}
+ T_1 + T_1' + T_2 + T_3
\right),
\end{equation}
where, using~(\ref{eq:extract_vals})
\begin{subequations}
    \begin{gather}
T_1 = \frac{3\lambda_0\,\bar{u}^2}{h_0^2}\,\left\langle \hat{v}\middle| \hat{m}^{(1)} \right\rangle, \hspace{0.3cm} T_1' = - \frac{3\lambda_0^2\,\bar{u}}{h_0^2}\,\left\langle \hat{v}\middle| \hat{m}^{(1)} \right\rangle^2,
\\
    T_2 = \sum_{i=0}^{N_q-2} v_i^3
\left(\frac{1}{h_i^2} - \frac{1}{h_{i+1}^2}\right)
+ \frac{v_{N_q-1}^3}{h_{N_q-1}^2}, \hspace{0.3cm}
T_3 = \sum_{i=0}^{N_q-2}
\frac{3}{h_{i+1}^2}
\left( v_i^2 v_{i+1} - v_i v_{i+1}^2 \right).
\end{gather}
\end{subequations}
After some algebra, the latter two terms take the forms 
\begin{gather}
T_2
= \lambda_0^3\,\|\boldsymbol{m}^{(4)}\|
\left\langle \hat{v}\middle|
D_{\hat{m}^{(4)}}\,D_{\hat{v}}
\middle| \hat{v} \right\rangle, \\
T_3
= 3\,\lambda_0^3\,\|\boldsymbol{m}^{(5)}\|
\left\langle \hat{v}\middle|
D_{\hat{m}^{(5)}}\,D_{\hat{v}}\,\widehat{A}
\middle| \hat{v} \right\rangle - 3\,\lambda_0^3\,\|\boldsymbol{m}^{(6)}\|
\left\langle \hat{v}\middle|
D_{\hat{m}^{(6)}}\,D_{\hat{v}}\,\widehat{A}^\dagger
\middle| \hat{v} \right\rangle,
\end{gather}
where
\begin{equation}
m^{(4)}_i =
\begin{cases}
\dfrac{1}{h_i^2} - \dfrac{1}{h_{i+1}^2}, & i = 0,\ldots,N_q-2, \\[6pt]
\dfrac{1}{h_{N_q-1}^2}, & i = N_q-1,
\end{cases}
\end{equation}
and

\begin{equation}
m^{(5)}_i =
\begin{cases}
\dfrac{1}{h_{i+1}^2}, & i = 0,\ldots,N_q-2, \\
0, & i = N_q - 1,
\end{cases}
\qquad
m^{(6)}_i =
\begin{cases}
0, & i = 0, \\
\dfrac{1}{h_{i}^2}, & i = 1,\ldots, N_q-1,
\end{cases}
\end{equation}
where the vanishing entries $m^{(5)}_{N_q-1}=0$ and $m^{(6)}_0=0$ suppress the periodic wrap-around terms $v_{N_q-1}^2v_0$ and $v_0^2v_{N_q-1}$ that the cyclic actions of $\widehat{A}$ and $\widehat{A}^\dagger$ would otherwise generate, respectively, so that $T_3$ reduces to the intended nonperiodic nearest-neighbor coupling.

Higher-order terms can be derived analogously and are omitted here for brevity.

The \emph{body force term} \eqref{eq:body_energy} is expressed as
\begin{equation}
\tilde{\mathcal{E}}_B(\boldsymbol{u})
=
- \left(b_0\,\bar{u}
+
\sum_{i=0}^{N_q-1} b_{i+1}\,v_i\right) ,
\end{equation}
where the coefficients $b_i$ are given by
\begin{equation}
b_i
=
\sum_{e=0}^{N_q-1}
\sum_{g=0}^{1}
\frac{h_e}{2}\,
B\!\left(X_e^{(g)}\right)\,
N_i\!\left(X_e^{(g)}\right),
\end{equation}
where 
$N_i\!\left(X_e^{(g)}\right)=0$
for elements not connected to node $i$.
Introducing the vector of interior force coefficients
\begin{equation}
\boldsymbol{b} := (b_1,b_2,\ldots,b_{N_q})
= \|\boldsymbol{b}\|\,\ket{\hat b},
\end{equation}
and recalling that $u_0=\bar{u}$ is prescribed, we obtain
\begin{equation}
\tilde{\mathcal{E}}_B(\boldsymbol{u})
=
-\left(b_0\,\bar{u}
+
\lambda_0\,\|\boldsymbol{b}\|\,
\left\langle \hat b \middle| \hat v \right\rangle\right),
\end{equation}
where the quantum state
\begin{equation}
\ket{\hat b} = \widehat{V}(\boldsymbol{\gamma})\ket{0}
\end{equation}
is prepared using the state-preparation procedure described in Section~\ref{sec:state_prep}.

The boundary contribution from \eqref{eq:energy_upto_cub} is given by
\begin{equation}
\label{eq_boundary_contribute}
-\bar{P}\,u(L)
=
-\lambda_0\,\|\boldsymbol{m}^{(-1)}\| \,\bar{P}\,
\left\langle \hat{v}\middle| \hat{m}^{(-1)} \right\rangle,
\end{equation}
where
\begin{equation}
\boldsymbol{m}^{(-1)}
:=
\|\boldsymbol{m}^{(-1)}\|\,\ket{\hat{m}^{(-1)}}
=
[0,0,\ldots,0,1].
\end{equation}

The expressions corresponding to the penalty-based formulation for the IHT-expansion-based approximation of $\mathcal{E}$ (cf.~Appendix~\ref{app:penalty_formulation}) are derived analogously.

In the above derivations, all expressions—such as Gauss-point interpolations and their powers—are explicitly expanded in terms of the global degree-of-freedom vector. This motivates the term \emph{direct-expansion} scheme. However, as the polynomial order or the order of the shape functions increases, the number of required quantum circuits grows rapidly. An alternative approach is to employ block encodings.

\subsubsection{A Block-encoding-based method}
\label{sec:block_encoding_2nd_order_formulas}
In this approach, the interpolation of nodal data to Gauss points, being a linear and generally non-unitary operation, is implemented via a block encoding of the interpolation matrix, as described in Section~\ref{sec:block_encode}. We demonstrate this approach for the second-order finite-element discretization of the energy functional presented in Section~\ref{sec:second_order_discretizzation}. Consider \eqref{eq:SF2_energy_terms} with a two-point Gauss quadrature scheme and a Dirichlet boundary condition at the left boundary. Rewriting \eqref{eq:SF2_cub_energy} by separating the contribution of the first element:
\begin{equation}
\tilde{\mathcal{E}}_k(\boldsymbol{u})
=
\mu\left(\frac{(-1)^k}{k}\right)
\sum_{g=0}^{1}
\left[
\left(\frac{2}{h_0}\right)^{k-1}
\left(\sum_{j=0}^{2} u_{j_0}\,\beta_j^{(II,g)}\right)^k
+
\sum_{e=1}^{N_q/2 - 1}
\left(\frac{2}{h_e}\right)^{k-1}
\left(\sum_{j=0}^{2} u_{j_e}\,\beta_j^{(II,g)}\right)^k
\right].
\label{eq:SF2_BE_formula}
\end{equation}
The contribution from the left boundary element can be constructed by finding the values of $u_1$ and $u_2$ (or equivalently $v_0$ and $v_1$) by considering their dot products, with appropriately prepared basis state vectors as considered in \eqref{eq:extract_vals}. 

To compute the (bulk) contribution of internal elements (i.e., the second term on the right-hand side of (\ref{eq:SF2_BE_formula})), consider the circulant coefficient matrix $\boldsymbol{\beta}^{(II,g)}$ as defined in \eqref{eq_circulant_matrix_II} and $\boldsymbol{\ell}_v^{(II,g)}$ using \eqref{eq:Block_encode_action_q}. Noticing that  \begin{equation}
\left(\boldsymbol{\ell}_v^{(II,g)}\right)_i =
\begin{cases}
\beta_0^{(II,g)}\, \hat{v}_i + \beta_1^{(II,g)}\, \hat{v}_{i+1} + \beta_2^{(II,g)}\, \hat{v}_{i+2}, & i = 0,\ldots,N_q-3, \\[6pt]
\beta_0^{(II,g)}\, \hat{v}_{N_q-2} + \beta_1^{(II,g)}\, \hat{v}_{N_q-1} + \beta_2^{(II,g)}\, \hat{v}_{0}, & i = N_q-2, \\[6pt]
\beta_0^{(II,g)}\, \hat{v}_{N_q-1} + \beta_1^{(II,g)}\, \hat{v}_{0} + \beta_2^{(II,g)}\, \hat{v}_{1}, & i = N_q-1,
\end{cases}
\end{equation}
the physically relevant, per-element contributions occupy only the odd-indexed entries $i=1,3,\ldots,N_q-3$. The even-indexed entries arise from spurious cross-element mixing inherent to the circulant structure of $\boldsymbol{\beta}^{(II,g)}$, while the entry at $i=N_q-2$ and $i=N_q-1$  arises from the periodic identification and correspond to a fictitious element beyond the last physical one; both must therefore be suppressed. Accordingly, we define the vector $\bbm^{(7)} \in \mathbb{R}^{N_q}$ such that
\begin{equation}
m^{(7)}_i =
\begin{cases}
\left(\dfrac{2}{h_{\frac{i+1}{2}}}\right)^{k-1}, & i = 1,3,\ldots,N_q-3, \\[6pt]
0, & \text{otherwise}.
\end{cases}
\label{eq:masking_in_BE}
\end{equation}
Following \eqref{eq:cubic_block_encode_formula}, the bulk contribution is then written as
\begin{equation}
\sum_{e=1}^{N_q/2 - 1}
\left(\frac{2}{h_e}\right)^{k-1}
\left(\sum_{j=0}^{2} u_{j_e}\,\beta_j^{(II,g)}\right)^k
= \lambda_0^k\, \|\bbm^{(7)}\|
 \left\langle \hat{v}  \,\middle|U^\dagger_{\boldsymbol{\beta^{(II,g)}}} \, D^{k-2}_{\bbll_v^{(II,g)}} \,D_{\bbm^{(7)} }\, U_{\boldsymbol{\beta^{(II,g)}}} \middle|\,  \hat{v} \,\right\rangle
\end{equation}
where the Hadamard product is applied $k - 1$ times. 
The quadratic term \eqref{eq:SF2_quad_energy} can be evaluated analogously.

To evaluate the \emph{body force} term \eqref{eq:SF2_body_energy}, we define
\begin{equation}
m^{(8,g)}_i =
\begin{cases}
\displaystyle \frac{1}{2}\,{h_{\frac{i+1}{2}}\, B\left(X_{\frac{i+1}{2}}^{(g)}\right)}, & i = 1,3,\ldots,N_q-3, \\[6pt]
0, & \text{otherwise},
\end{cases}
\end{equation}
where the masking follows the same logic as employed  for \eqref{eq:masking_in_BE}.
Then, using \eqref{eq_circulant_matrix_II} and \eqref{eq:Block_encode_action_q}, the bulk contribution becomes
\begin{equation}
\sum_{e=1}^{N_q/2 - 1}
\frac{h_e}{2}
\left(\sum_{j=0}^{2} u_{j_e}\,\beta_j^{(I,g)}\right)
B(X_e^{(g)})
 = \lambda_0\,\|\boldsymbol{m}^{(8,g)}\|
\left\langle \hat{m}^{(8,g)}\middle|
U_{\boldsymbol{\beta^{(I,g)}}}
\middle| \hat{v} \right\rangle.
\end{equation}
The contribution of the first element to the body force term is obtained by substituting the values of \(u_1\) and \(u_2\), determined in the treatment of \eqref{eq:SF2_cub_energy}, into \eqref{eq:SF2_body_energy} and evaluating the resulting expression (contribution from first element only) classically.

\section{Complexity}
\label{sec:complexity} 

The complexity of the proposed algorithm is computed in terms of cost per quantum circuit evaluation and the number of evaluations required until convergence is achieved in classical optimization. The cost of a given circuit is estimated in terms of its depth, which is evaluated by decomposing the circuit into a basis gate set that consists of one- and two-qubit operations (i.e., Hadamard, Pauli, phase, rotation and controlled rotation gates). We utilize \emph{Qiskit Optimization 3} ~\cite{qiskit2024} to perform the circuit decompositions. 

\begin{figure}[t]
    \centering
    \includegraphics[width=0.6\columnwidth]{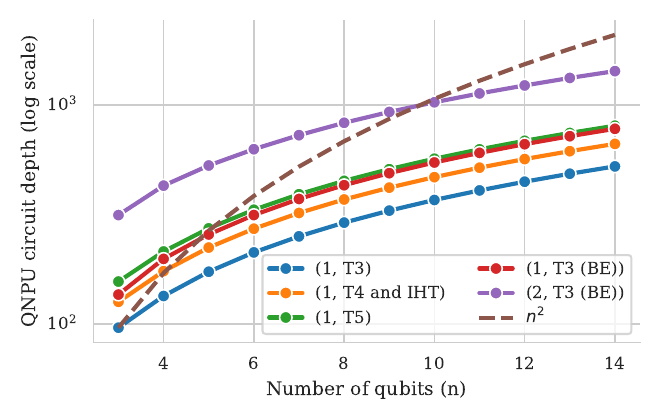}
    \caption{Scaling of QNPU circuit depth (including block-encoding where applicable) with the number of qubits $n$ for different formulations. The depth is shown on a logarithmic scale; the dashed line indicates quadratic scaling $\mathcal{O}(n^2)$. Legend entries denote (shape function order, expansion method).}
    \label{fig:qnpu_depth}
\end{figure}

As per the QNPU framework (see Fig.~\ref{fig:VQA_outline}), the total depth (excluding the Hadamard test for simplicity) of a circuit is given by 
\begin{equation}
D_c = D_{\text{input}} + D_{\text{QNPU}},
\end{equation}
where $D_{\text{input}}$ is the depth required by the preparation step (shown as green in Fig.~\ref{fig:VQA_outline}) and $D_{\text{QNPU}}$ is the depth of the QNPU. If block encoding is used, an additional term $D_{\text{BE}}$ is also included (see Fig.~\ref{fig:block_encode_qnpu}). The depth required by the preparation step is given by:
\begin{equation}
    \label{eq:ansatz_input_depth}
    D_{\text{input}}=\mathcal{O}(D_{\text{ansatz}}),
\end{equation}
where $D_{\text{ansatz}}$ indicates the depth required to produce one of the controlled input ports and the $D_{\text{input}}$ is a small multiple of $D_{\text{ansatz}}$ depending on the number of input ports in the circuit. $D_{\text{ansatz}}$ itself is dependent on the entanglement structure of the ansatz, the number of repeating layers  $d$ and the overhead to implement the ansatz in a controlled way. The scaling of $D_{\text{ansatz}}$ is
$D_{\text{ansatz}} = \mathcal{O}(nd)$. The contributions from $D_{\text{QNPU}}$ (and $D_{\text{BE}}$) are independent of the ansatz structure and their scaling in $n$ is shown in Fig.~\ref{fig:qnpu_depth}. The figure indicates $\mathcal{O}(n)$ scaling. Thus we obtain 
\begin{equation}
    D_c = \mathcal{O}(nd).
\end{equation}

The total complexity of the algorithm is then expressed as
\begin{equation}
\mathcal{C}_{\mathrm{tot}} = T_{\mathrm{it}} \, N_{\mathrm{eval/it}} \, N_{\mathrm{circ/eval}} \, s \, D_c,
\end{equation}
where $T_{\mathrm{it}}$ denotes the number of classical optimizer iterations, $N_{\mathrm{eval/it}}$ the number of cost-function evaluations per iteration, $N_{\mathrm{circ/eval}}$ the number of circuits evaluated per cost-function call, and $s$ the number of measurement shots per circuit.

For gradient-based methods, $N_{\mathrm{eval/it}}$ typically scales with the number of variational parameters \cite{Schuld2019}, i.e., $\mathcal{O}(p)$ (or equivalently $\mathcal{O}(nd)$). The number of circuits per evaluation, $N_{\mathrm{circ/eval}}$, depends on the specific formulation but is independent of  $n$. Thus,
\begin{equation}
N_{\mathrm{circ/eval}} = \mathcal{O}(1).
\end{equation}

The number of measurement shots per circuit depends on the desired accuracy $\epsilon$ of the estimated expectation values. In the direct-expansion scheme, the increasing number of circuits may require higher precision per circuit, thereby increasing the shot count. In the block-encoding approach, the number of circuits may be small, however, additional challenges may arise from the post-processing of ancilla measurements. 
In the absence of noise, $s$ does not scale with $n$ or circuit complexity. Therefore, shot estimation is excluded from the complexity analysis.

In addition to the contributions mentioned above, the algorithm requires fixed control parameters to prepare several auxiliary states. The number of such states scales as $\mathcal{O}(1)$ in terms of $n$. A classical pre-optimization step is employed to determine the parameters required for each auxiliary state. The subsequent preparation of each auxiliary state is included in $D_{\text{input}}$, as defined in \eqref{eq:ansatz_input_depth}, for every circuit in which that state is required. When combined, the total complexity in $n$ yields
\begin{equation}
\label{eq:complexity}
\mathcal{C}_{\mathrm{tot}} = \mathcal{O}\Bigl( T_{\mathrm{it}} \cdot (nd) \cdot N_{\mathrm{circ/eval}} \cdot (nd) \Bigr) = \mathcal{O}\Bigl( T_{\mathrm{it}} n^2 d^2 \Bigr).
\end{equation}
Comparing against a classical approach where the nonlinear problem is solved using an iterative approach applied to solving a system of equations at each step: the classical cost can be approximated as 
\begin{equation}
\mathcal{C}_{cl} = \mathcal{O}(T_{cl} \cdot N_q); \quad N_q \sim 2^n,
\end{equation}
where the Thomas algorithm for tridiagonal systems has a cost of $\mathcal{O} (N_q)$ at each iteration and $T_{cl}$ represents the number of classical iterations. The proposed complexity \eqref{eq:complexity} may achieve a polylogarithmic scaling in $N_q$ only if $d$, and $T_{it}$ remain polynomial in $n$, and efficient state preparation techniques can be employed, potentially leading to an exponential asymptotic improvement over classical iterative solvers. The practical realization of this scaling and the achievement of an effective quantum gain, however, are hindered by several algorithmic and hardware-level issues, which are discussed in Sections~\ref{sec:challenges} and \ref{sec:conclude}.

\section{Numerical Verification}
\label{sec:examples}
The proposed approach has been implemented in Python. The cost function and Jacobian computations were performed via the quantum algorithm by employing Qiskit 2.2.3. State Vector Simulator method provided by Backend AER without noise was used. The performance and capabilities of the proposed approach have been assessed using a series of numerical experiments discussed below. 

BFGS (Broyden–Fletcher–Goldfarb–Shanno) optimization method available within the SciPy Optimization toolkit was used. The cost function gradients were evaluated using finite differences set at default values of $\sqrt{\epsilon_{\text{machine}}}$ \cite{Fletcher1970, NocedalWright}. The optimization is terminated when either of the following {criterion} is satisfied:
\begin{equation}
\label{eq:stopping_criterion}
\left| \mathcal{C}^{(k)} - \mathcal{C}^{(k-1)} \right|
<
\begin{cases}
10^{-7}, & \text{for three consecutive iterations},\\
10^{-8}, & \text{for a single iteration},
\end{cases}
\end{equation}
where $\mathcal{C}^{(k)}$ denotes the value of the cost function \eqref{eq:cost_cum_energy} at iteration $k$.
The effect of the stopping criterion is demonstrated using an example in Section~\ref{sec:EX_P2}. 

The initial guess for the control parameters used in the cost-function optimization is selected such that the resulting displacement profile satisfies the constraint \eqref{eq:diap_grad_constraint}. Accordingly, random displacement profiles are generated such that $-\alpha < u' < \alpha$, $\alpha = 0.5$. Control parameters for a profile that satisfies this constraint are created using the state-preparation procedure described in Section~\ref{sec:state_prep}. For the IHT-based formulation, the initial guess for the $u$ profile (and the associated control parameters) was generated using the same procedure, while the $y$ profile (and its corresponding control parameters) was obtained using \eqref{eq:y_f(u)}.  

The errors in the computed solutions may arise from multiple sources, including the polynomial approximation of the nonlinear logarithmic term in~\eqref{eq:strong_form_neo}, the penalty formulation employed for the IHT approximation, finite-element discretization, and the VQA solution procedure. The convergence of the polynomial and penalty approximations is analyzed separately in Appendix~\ref{app:constitutive_error}. 

Here, we introduce the error measures used to assess the overall solution accuracy and to distinguish the VQA-related error from the other contributions. In particular, the algorithm computes the solution corresponding to the approximated strain energy density function instead of the original hyperelastic model. Accordingly, we report the maximum deviation in displacement gradient and the $L_2$ displacement error (relative to the analytical solution of \eqref{eq:strong_form_neo} $u^{\mathrm{NLS}}$ -- percentage measure):
\begin{equation}
E_{|u'|}^{\mathrm{NLS}} 
= 
\max_{X \in \Omega_0}\,\bigl|u^{\mathrm{VQ}^{'}}(X) - u^{\,\mathrm{NLS}^{'}}(X)\bigr|,
\qquad
E_{L_2}^{\mathrm{NLS}}
=
\frac{
\left\|u^{\mathrm{VQ}}-u^{\mathrm{NLS}}\right\|_{L_2(\Omega_0)}
}{
\left\|u^{\mathrm{NLS}}\right\|_{L_2(\Omega_0)}
}
\times 100,
\end{equation}
where $u^{VQ}$ represents the FE field for displacement obtained using the VQA approach and 
\begin{equation}
\left\|u\right\|_{L_2(\Omega_0)}
=
\left(
\int_{\Omega_0}
\left|u(X)\right|^2
\,\mathrm{d}X
\right)^{1/2}.
\end{equation}
 To isolate the error arising due to the introduction of VQA, we measure the fidelity of the quantum state obtained from the VQA relative to a reference state, generated by minimizing the same energy functional classically, directly for the displacement field. We refer to the classically obtained  reference profile as $u^{\text{Cl}}$, the corresponding quantum state as $
\ket{\hat{u}^{\mathrm{Cl}}}
$ and compute the fidelity as:
\begin{equation}
E_{\mathrm{tr}}^{\mathrm{Cl}}
=
\sqrt{
1 -
\left|
\left\langle
\hat{u}^{\mathrm{VQ}}
\middle|
\hat{u}^{\mathrm{Cl}}
\right\rangle
\right|^2
},
\end{equation}
where $\ket{\hat{u}^{\mathrm{VQ}}}$ represents the quantum state obtained using the optimal quantum parameters based on the chosen ansatz.

All errors are evaluated over a sample of 20 successful independent runs per example. In each run, both the classical optimization and the VQA-based optimization are initialized using the same initial guess for the classical DoFs $\bbu$ (and auxiliary variables $\bby$, where applicable), chosen within the admissible bounds. The reported values correspond to averages over these runs.

\subsection{Effect of nonlinearity approximation}

In this section, the accuracy and computational costs of the proposed method are investigated relative to the polynomial approximation of the logarithmic term in the Neo-Hookean strain energy density function. Consider a domain of unit length ($L=1$ mm) discretized into 8 first-order (i.e., linear) elements (with 8 DoFs excluding the Dirichlet boundary node) using a uniform finite element grid. The domain is subjected to a linear body force \begin{equation}
\label{eq:body_force}
    B(X) = \kappa X,
\end{equation} and traction-free boundary at $X=L$ (i.e., $\bar{P}=0$ N/mm$^2$). The modulus and the body force constants are set to $\mu = 1$ MPa and $\kappa = 1.5$ N/mm$^4$. The direct-expansion-based method from Section \ref{sec:Direct_expansion} has been employed to obtain the results.

\begin{figure}[t]
    \centering

    \begin{subfigure}[t]{0.48\textwidth}
        \centering
        \includegraphics[width=\linewidth]{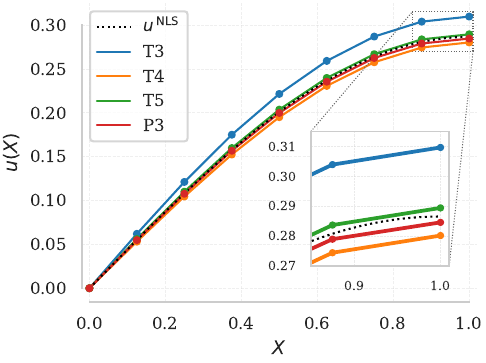}
        \caption{}
        \label{fig:EX1_profiles}
    \end{subfigure}
    \hfill
    \begin{subfigure}[t]{0.48\textwidth}
        \centering
        \includegraphics[width=\linewidth]{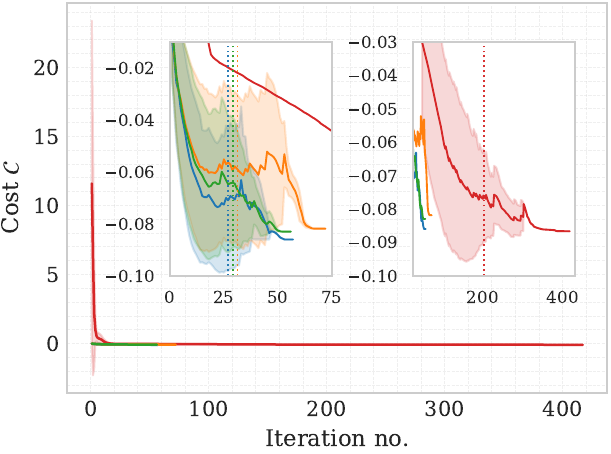}
        \caption{}
        \label{fig:EX1_iteration}
    \end{subfigure}

    \vspace{0.4cm}

    \begin{subtable}[t]{\textwidth}
        \centering
        \begin{tabular}{l c c c c}
        \toprule
         & Evaluations & $E_{L_2}^{\mathrm{NLS}} (\%)$ & $E_{|u'|}^{\mathrm{NLS}}$ & $E_{\mathrm{tr}}^{\mathrm{Cl}}$ \\
        \midrule
        T3 & $403 \pm 161$ & $8.92 \pm 2.41 \times 10^{-3}$ & $6.09 \times 10^{-2} \pm 2.89 \times 10^{-5}$ & $1.89 \times 10^{-5} \pm 1.36 \times 10^{-5}$ \\
        T4 & $444 \pm 208$ & $2.86 \pm 9.75 \times 10^{-3}$ & $4.59 \times 10^{-2} \pm 1.42 \times 10^{-5}$ & $2.07 \times 10^{-5} \pm 1.79 \times 10^{-5}$ \\
        T5 & $432 \pm 159$ & $0.92 \pm 7.90 \times 10^{-3}$ & $4.59 \times 10^{-2} \pm 5.91 \times 10^{-5}$ & $4.11 \times 10^{-5} \pm 1.57 \times 10^{-5}$ \\
        P3 & $4699 \pm 1635$ & $0.9 \pm 3.19 \times 10^{-2}$ & $4.62 \times 10^{-2} \pm 4.40 \times 10^{-3}$ & $6.13 \times 10^{-4} \pm 1.56 \times 10^{-3}$ \\
        \bottomrule
        \end{tabular}
        \caption{}
        \label{tab:EX1_results}
    \end{subtable}

    \caption{
    VQA performance (20-run average) for Taylor orders 3–5 (T3–T5) and IHT (P3). The profiles of $u^{VQ}$ (in mm) and the iteration cost histories are shown in (a) and (b), respectively. Performance metrics for the different models are summarized in Table~(c), where “Evaluations” denotes the average number of function calls over the entire optimization process.
    }
    \label{fig:EX1_results}
\end{figure}

Fig.~\ref{fig:EX1_results} shows the predicted displacement profiles and the evolution of the cost function as a function of optimization iterations for Taylor and IHT-expansion-based approximations. The internal DoFs are represented using a 3-qubit main register ($n = 3$). The ansatz layer depth is set to $d = 2$, resulting in 10 control parameters for the models that use Taylor approximations, and 20 (10 for displacement and 10 for auxiliary field) for the model that uses IHT approximation. Three models use Taylor approximations (T3, T4, T5 with three, four, and five-term expansions, respectively), whereas the model with IHT approximation (P3) employs an expansion up to and including the cubic term. {The total number of distinct circuits required for T3, T4, T5, and P3 are 7, 11, 16, and 21, respectively. The most expensive circuits for these cases require 11, 14, 17, and 14 qubits, respectively.} The P3 model uses a penalty parameter of $\mathcal{P} = 10^2$. 

   The displacement profiles in Fig.~\ref{fig:EX1_profiles}, together with the $L_2$ errors listed in Table~\ref{tab:EX1_results}, show that the numerical solution approaches the nonlinear analytical solution as the order of the polynomial approximation increases. The analytical solution to the problem is derived in Appendix~\ref{app:exact_sol_deriv}. {The small trace errors indicate that the computed profiles $u^{VQ}$ closely match the expected profiles that would be obtained using a classical minimizer applied to the approximated energy functional.} Among the considered methods, the IHT-based method yields the most accurate results. However, the IHT-based method exhibits convergence difficulties, reflected in larger standard deviations of the error metrics and slower convergence in Fig.~\ref{fig:EX1_iteration}, especially near the minimizer. These convergence difficulties are attributed to the penalty formulation, specifically the {ill-conditioning} introduced by the penalty parameter, $\mathcal{P}$. Setting small values to $\mathcal{P}$ leads to ineffective enforcement of the constraint, whereas large values can cause ill-conditioning, as is well known in optimization literature~\cite{NocedalWright}. 

\begin{figure}[htbp]
    \centering

    \begin{subfigure}[t]{0.48\textwidth}
        \centering
        \includegraphics[width=\linewidth]{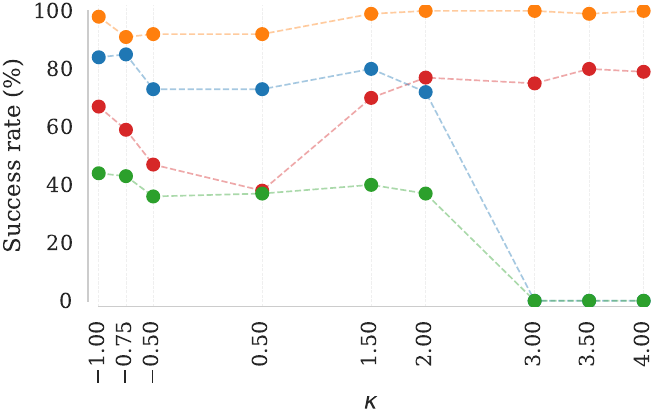}
        \caption{}
        \label{fig:success}
    \end{subfigure}
    \hfill
    \begin{subfigure}[t]{0.48\textwidth}
        \centering
        \includegraphics[width=\linewidth]{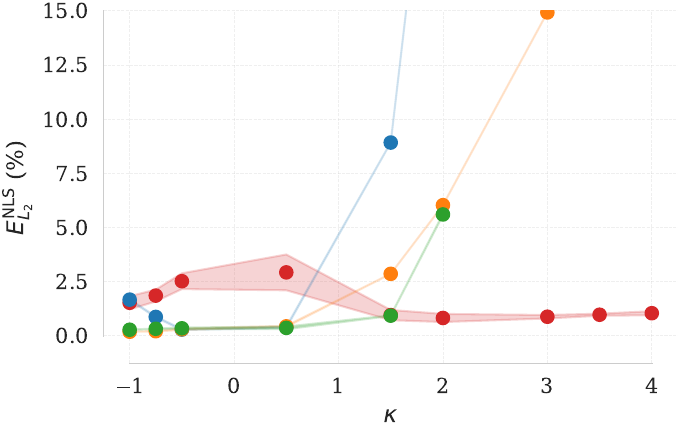}
        \caption{}
        \label{fig:l2}
    \end{subfigure}

    \caption{Performance of different models as a function of $\kappa$ (N/mm$^4$)(see \eqref{eq:body_force}), averaged over 100 runs: (a) success rate and (b) $L_2$ error relative to the analytical solution.}
    \label{fig:app_nonlinear}
\end{figure}

Fig.~\ref{fig:app_nonlinear} demonstrates that polynomial approximations can degrade the convexity of the potential energy functional, leading to divergence, convergence to nonphysical solutions, or optimization trajectories entering regions where $u' < -1$. To quantify this behavior, $\kappa$ in \eqref{eq:body_force} is varied and the corresponding success ratio (fraction of initial guesses that converge under the prescribed optimization criteria) and $E_{L_2}^{\mathrm{NLS}}$ are evaluated. The resulting success rates (computed over 100 runs) and errors (computed over 20 successful runs) are reported in Fig.~\ref{fig:success} and Fig.~\ref{fig:l2}, respectively. Among the Taylor expansions, the fourth-order approximation preserves convexity more effectively than the third- and fifth-order approximations, yielding consistently higher success rates. However, its accuracy deteriorates for larger $\alpha$, with $E_{L_2}^{\mathrm{NLS}}$ reaching approximately $15\%$ at $\kappa = 3$. Beyond $\kappa\geq3$, analytical solution demands $u'>1$, invalidating the Taylor-expansion-based models. In contrast, the third-order IHT expansion performs well away from $\alpha \approx 0$ but exhibits lower success rates and larger errors near this regime. Overall, Taylor expansions are better suited for small deformations, whereas the IHT-expansion-based method is more reliable for larger deformations.

\subsection{Effect of problem size} \label{sec:EX_P2}

In this section, the characteristics of the solution process as a function of the problem size are investigated. We report scaling results only for the Taylor-expansion-based method up to the third order. The direct-expansion-based method from Section \ref{sec:Direct_expansion} has been employed to obtain the results. Higher-order expansions require additional ancilla qubits that significantly increase the size of the state vector and the associated memory cost. This makes large problems difficult to implement on classical simulators. Although approaches have been recently proposed to alleviate these limitations to some extent~\cite{10.1145/3665283.3665293}, they are not considered in the present work.

\begin{figure}[t]
    \centering

    \begin{subfigure}[t]{0.49\textwidth}
        \centering
        \includegraphics[width=\linewidth]{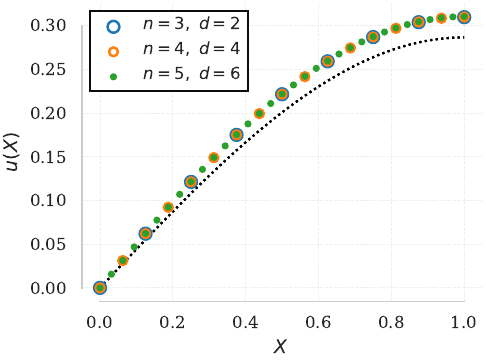}
        \caption{}
        \label{fig:P2_profiles}
    \end{subfigure}
    \hfill
    \begin{subfigure}[t]{0.49\textwidth}
        \centering
        \includegraphics[width=\linewidth]{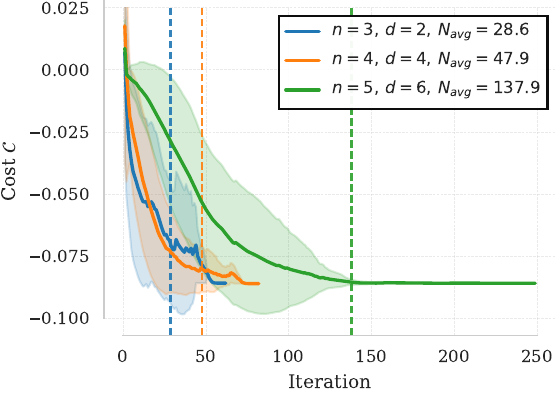}
        \caption{}
        \label{fig:P2_iterations}
    \end{subfigure}

    \vspace{0.4cm}

    \begin{subtable}[t]{\textwidth}
        \centering
        \begin{tabular}{lccccccc}
        \toprule
         $n$ & $d$ & $p$ & Avg. Func. Calls & 
         $E_{L_2}^{\mathrm{NLS}} (\%)$ & 
         $E_{|u'|}^{\mathrm{NLS}}$ & 
         $E_{\mathrm{tr}}^{\mathrm{Cl}}$ \\
        \midrule
         $3$ & $2$ & $10$  & $399 \pm 180$  & $8.9 \pm 1.9 \times 10^{-3}$ &
         $6.1 \times 10^{-2} \pm 6.2 \times 10^{-5}$ &
         $1.3 \times 10^{-5} \pm 7.7 \times 10^{-6}$ \\

         $3$ & $4$ & $16$ & $601 \pm 224$ &
         $8.9 \pm 2.3 \times 10^{-3}$ &
         $6.1 \times 10^{-2} \pm 4.8 \times 10^{-5}$ &
         $1.4 \times 10^{-5} \pm 1.5 \times 10^{-5}$ \\

         $4$ & $4$ & $21$ & $1277 \pm 373$ &
         $9.2 \pm 2.4 \times 10^{-2}$ &
         $5.8 \times 10^{-2} \pm 1.8 \times 10^{-4}$ &
         $3.5 \times 10^{-5} \pm 2.3 \times 10^{-5}$ \\

         $4$ & $6$ & $29$ & $1828 \pm 482$ &
         $9.2 \pm 1.5 \times 10^{-2}$ &
         $5.8 \times 10^{-2} \pm 1.1 \times 10^{-4}$ &
         $3.1 \times 10^{-5} \pm 2 \times 10^{-5}$ \\

         $5$ & $6$ & $36$ & $5731 \pm 1722$ &
         $9.4 \pm 8.3 \times 10^{-1}$ &
         $6.4 \times 10^{-2} \pm 1.1 \times 10^{-2}$ &
         $9.9 \times 10^{-4} \pm 9.7 \times 10^{-4}$ \\

         $5$ & $8$ & $46$ & $4761 \pm 1113$ &
         $9.3 \pm 10 \times 10^{-2}$ &
         $5.8 \times 10^{-2} \pm 6.6 \times 10^{-4}$ &
         $1.4 \times 10^{-4} \pm 5.6 \times 10^{-5}$ \\
        \bottomrule
        \end{tabular}
        \caption{}
        \label{tab:P2_results}
    \end{subtable}

    \caption{
    VQA scaling for $n=3,4,$ and $5$ qubits using a third-order Taylor expansion (20-run average). 
    The profiles of $u^{VQ}$ (in mm) and the iteration-cost histories are shown in (a) and (b), respectively. 
    Performance metrics are summarized in Table~(c). Vertical dotted lines in (b) indicate the average convergence iterations $N_{\text{avg}}$ for the minimum ansatz layer depth required by each value of $n$.
    }
    \label{fig:vqe_comparison}
\end{figure}

Fig.~\ref{fig:vqe_comparison} shows the results obtained for the problem setting discussed in the previous section, but with three increasingly refined meshes. Internal DoF encoded in the main qubit register are $n = 3, 4, 5$ with variational ansatz layer depths set to $d = \{2,4\}$, $\{4,6\}$, and $\{6,8\}$, respectively. The number of distinct circuits is independent of the number of qubits, whereas the size of the most expensive circuit scales as $4n - 1$. As mesh refinement increases the problem size and enlarges the displacement vector, a richer ansatz is required to adequately parameterize the solution space and to retain expressivity. The minimum depths selected for each value of $n$ are those required to achieve trace errors below $10^{-3}$. Increasing the circuit depth (and thus the number of parameters) does not significantly affect the metrics for $n=3$ and $n=4$. However, for $n=5$, improvements across all reported metrics (see Table~\ref{tab:P2_results}) indicate that systems with a larger number of qubits may require additional parameters to achieve better performance. We have further investigated the relationship between ansatz layer depth and the number of qubits in the main register as well as the polynomial approximation order. This study is shown in Appendix \ref{sec:expressivity}, where we observe no sensitivity to polynomial approximation order and relatively mild dependence on $n$ within the range of problems investigated. 

Fig.~\ref{fig:P2_profiles} shows that the displacement profiles obtained using the VQA approach converge to the same global solution upon refinement, while Fig.~\ref{fig:P2_iterations} presents the corresponding optimization histories. 
Increasing the problem size  introduces convergence difficulties near the minimizer.  
This is evident from the long tail  noticeable in the optimization history for $n=5$. In contrast, smaller systems converge within a reasonable precision tolerance.

\subsection{Assessment of robustness}
In this section, several additional aspects of the model implementation are evaluated. We consider the following setup: The problem domain is discretized using two non-uniform meshes. The first mesh consists of eight linear finite elements with element lengths (in mm) of $[0.25,\,0.25,$ $0.1,\,0.1,$ $0.05,\,0.05,$ $0.1,\,0.1]$. The second mesh consists of four quadratic finite elements with element lengths of $[0.5,\,0.2,$ $0.1,\,0.2]$. The internal DoFs for both meshes are encoded in a 3-qubit main register ($n = 3$). The domain is subjected to a quadratic body force (i.e., $B(X) = 5X^2$ N/mm$^3$, $X$ in mm) with a prescribed traction at $X=L$ ($\bar{P}=-0.8$ N/mm$^2$) and a non-homogeneous Dirichlet condition at $X=0$ ($\bar{u}=0.1$ mm). The modulus of the domain is set to $\mu=1$ MPa. 
The logarithmic nonlinearity in the strain energy density is approximated using a three-term Taylor series approximation. In case of the mesh with linear elements, both direct expansion (D1) and Block encoding-based (BE1) approaches (Section \ref{sec:block_encoding_2nd_order_formulas}) were used to establish the circuits associated with the resulting cost function. In the case of the mesh with quadratic elements, the Block encoding (BE2) scheme was used.  

\begin{figure*}[t]
    \centering

    \begin{subfigure}[t]{0.49\textwidth}
        \centering
        \includegraphics[width=\linewidth]{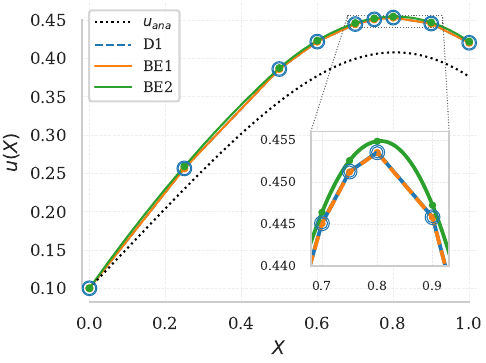}
        \caption{}
        \label{fig:P3_profiles}
    \end{subfigure}
    \hfill
    \begin{subfigure}[t]{0.49\textwidth}
        \centering
        \includegraphics[width=\linewidth]{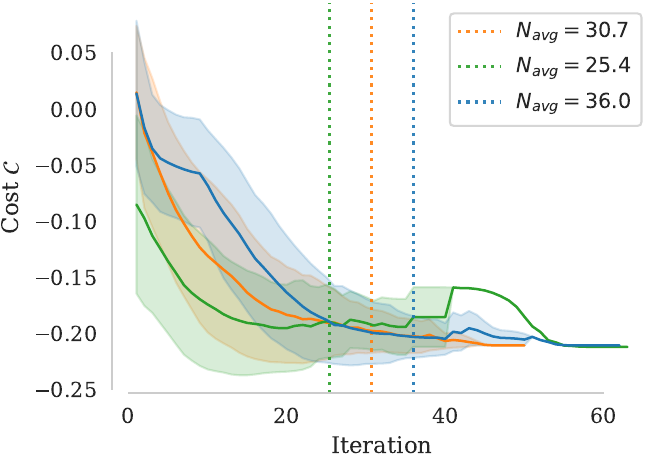}
        \caption{}
        \label{fig:P3_iterations}
    \end{subfigure}

    \vspace{0.4cm}

    \begin{subtable}[t]{\textwidth}
        \centering
        \begin{tabular}{cccccc}
        \toprule
        Formulation & Evaluations & 
        $E_{L_2}^{\mathrm{NLS}}$ (\%) & 
        $E_{|u'|}^{\mathrm{NLS}}$ & 
        $E_{\mathrm{tr}}^{\mathrm{Cl}}$ \\
        \midrule

        BE1 & $454 \pm 115$ &
        $11 \pm 9.8 \times 10^{-3}$ &
        $0.14 \pm 1.1 \times 10^{-4}$ &
        $2.4 \times 10^{-5} \pm 1.3 \times 10^{-5}$ \\

        BE2 & $402 \pm 149$ &
        $12 \pm 2 \times 10^{-2}$ &
        $0.13 \pm 2.2 \times 10^{-4}$ &
        $9.6 \times 10^{-5} \pm 4.4 \times 10^{-5}$ \\

        D1 & $652 \pm 148$ &
        $11 \pm 1.6 \times 10^{-2}$ &
        $0.14 \pm 1.2 \times 10^{-4}$ &
        $3.6 \times 10^{-5} \pm 2.8 \times 10^{-5}$ \\

        \bottomrule
        \end{tabular}
        \caption{}
        \label{tab:P3_results}
    \end{subtable}

    \caption{
    VQA performance comparison for different formulations (20-run average). The formulation notation denotes BE1 and BE2 as the block-encoding-based approaches employing first- and second-order shape functions, respectively, while D1 denotes the direct-expansion-based approach employing first-order shape functions. 
    The profiles of $u^{VQ}$ (in mm) and the iteration-cost histories are shown in (a) and (b), respectively. Vertical dotted lines in (b) indicate the average convergence iterations $N_{\text{avg}}$.
    Performance metrics are summarized in Table~(c).  
    }
    \label{fig:P3_figures}
\end{figure*}

Fig.~\ref{fig:P3_figures} presents the displacement fields and optimization histories obtained using the proposed approach. The two linear discretization models exhibit comparable convergence behavior and error characteristics, as summarized in Table~\ref{tab:P3_results}. In contrast, the second-order shape function model yields larger errors, which may be attributed to the use of only two Gauss integration points. Compared to the analytical solution, the errors observed in this example are higher than those reported in the preceding cases, as the present problem was intentionally selected to assess the limitations of the proposed scheme. Nevertheless, the low trace errors across all formulations indicate accurate solution recovery by the VQA approach.

From a structural perspective, the block-encoding approaches (BE1 and BE2) require only 3 circuits compared to 8 for the direct approach, albeit with additional ancilla and postselection overheads. The largest BE1 and BE2 circuits require 15 and 17 qubits, with 4 and 6 ancillas requiring postselection measurement, respectively (Section~\ref{sec:block_encode}), posing a challenge to their implementation. Similar to the example in Section~\ref{sec:EX_P2}, upon relaxing the optimization criterion in Eq.~\eqref{eq:stopping_criterion}, it was observed that the non-uniform mesh configuration requires substantially higher cost-function precision before the gradient norm falls below the prescribed tolerance. 

\section{Practical Considerations and Future Directions}
\label{sec:challenges}
\paragraph{State Preparation for Auxiliary vectors:} Preparation of arbitrary quantum states is a well-known bottleneck in quantum computing. The complexity to prepare an arbitrary state scales exponentially with number of qubits \cite{Mottonen2005StatePreparation}.
 For states with known structure, such as sparse vectors or amplitudes described by low-degree polynomial functions, more efficient structured techniques may be used \cite{PhysRevA.110.032609,OBrien2025quantumstate}. This may apply, for example, to analytically defined body-force fields. Similarly, for a uniform mesh, several auxiliary vectors $\bbm$ used in this work reduce to uniform superpositions $\frac{1}{\sqrt{M}}\sum_{j\in S}\ket{j}$ over a known index set $S$ with $|S|=M$, enabling ancilla-free preparation with depth $\mathcal{O}(\log_2 M)$ \cite{Shukla2024uniform} and thereby bypassing the optimization-based procedure in Eq.~\eqref{eq:state_prep}. Hence, based on the structure of the problems, one may reduce the state preparation overhead. 

\paragraph{Number of circuits and extension to multi-dimensional problems}
In the direct-expansion-based method, for a displacement approximation employing shape functions of order $n_{s}$, the number of circuit evaluations, $N_{\mathrm{circ/eval}}$, can be shown to scale as $\mathcal{O}(N^{n_{s}+1})$ for the Taylor-expansion-based method and as $\mathcal{O}(N^{n_{s}})$ for the IHT-expansion-based method, where $N$ denotes the number of terms retained in the respective expansions. If higher-order approximations are required for improved accuracy, the number of circuit evaluations can grow rapidly, while larger problem sizes may also incur substantial ancilla overheads. These estimates characterize the complexity associated with the nonlinear expansions considered in the present work; extension to higher-dimensional problems will introduce additional considerations, including multiple displacement components, coupled strain measures, increased degrees of freedom, and more complex nonlinearities. The present 1D study establishes the feasibility of the proposed formulations in a controlled proof-of-concept setting, while their computational scalability and efficient extension to realistic 2D and 3D hyperelastic problems remain subjects for future investigation.

To alleviate some of these issues, alternative formulations may be considered. For example, to handle the logarithmic term appearing in \eqref{eq:energy_func}, we introduce an auxiliary variable $z$ defined as
\begin{equation}
z := \ln(1+u')
\quad \Rightarrow \quad
c(u,z) := u'' - (1+u')z' = 0.
\end{equation}
The resulting relation can be imposed through a quadratic penalty, leading to the modified energy functional
\begin{equation}
\label{eq:log_form_2}
\mathcal{E}[u,z] = \int_{\Omega_0} \left[
\mu \left( u' + \tfrac{1}{2}(u')^2 - z \right)
- B\,u
+ \frac{\mathcal{P}}{2} c(u,z)^2
\right]\,\mathrm{d}X
- \bar{P}\,u(L).
\end{equation}
For a sufficiently large penalty parameter, $\mathcal{P} \gg 1$, minimizing $\mathcal{E}[u,z]$ approximately enforces $c(u,z)=0$, which recovers $z=\ln(1+u')$ up to an integration constant. With an appropriate additional condition on $z$ to fix this constant, the constrained formulation becomes consistent with the original energy functional in \eqref{eq:energy_func}. This reformulation eliminates the need to approximate the logarithmic term and, at the same time, expresses all terms in the integrand in terms of polynomial nonlinearities. Similarly, an iterative functional approach has been developed in \cite{kouskiya2026hyperelasticity}, in which an approximate low-order solution is first obtained and subsequently used as a reference state for expanding the nonlinear terms. This procedure generates a sequence of iterative VQAs, where each successive problem solves for a correction to the preceding approximation, thereby progressively improving the solution accuracy. The development of such alternative approaches is expected to become increasingly important when extending the current framework to higher-dimensional problems.

In contrast to the direct-expansion-based approach, the block-encoding-based approach can significantly reduce the number of circuit growth overhead, as element-wise arithmetic operations are effectively handled within the block-encoding framework.  However, the post-selection success probability can increase the measurement overhead significantly.  While evaluating a circuit involving the computation for $k^{\text{th}}$ order of $u'$, the success probability will reduce as $p_s^{k-1}$, where $p_s$ is the success probability of a single block-encoding. Even in the absence of hardware noise, implementation on quantum hardware would require a large number of measurement shots. Amplitude amplification could help reduce this overhead \cite{brassard2002amplitude}. Hence, for practical considerations, efficient block-encoding methods to generate the effect of block-encoding across copies of multiple states within the same circuit will need to be explored.  

\section{Conclusion}
\label{sec:conclude}

This manuscript presented a variational quantum algorithm (VQA)–based framework for solving nonlinear elasticity problems arising from hyperelastic material models. The proposed approach approximates nonlinearities in the strain energy density using polynomial expansions, enabling their implementation within the hybrid quantum-classical framework via Quantum Nonlinear Processing Units (QNPUs). The methodology was demonstrated on a special case of one-dimensional Neo-Hookean problem, where the potential energy functional is used as the VQA cost function. Two types of polynomial approximations were investigated, and the VQA was shown to capture the expected behavior using both first- and second-order finite element discretizations on non-uniform meshes with non-homogeneous boundary conditions. A few other types of nonlinearities have been explored in \cite{kouskiya2026hyperelasticity}.

The study highlights several practical challenges, including the influence of polynomial approximations on accuracy, convergence difficulties in more complex formulations, and increased computational cost associated with higher-order expansions and auxiliary variables. In addition, current implementations are limited by the cost of statevector simulations, particularly for larger problem sizes, restricting the analysis of scaling behavior.
The impact of hardware noise, coupling latency, native-gate-set compilation, and cloud-access queue times on the performance of the framework when deployed on real hardware requires further optimization of algorithmic choices and efficient hardware-aware implementation.
Future work will focus on addressing these limitations. In particular, developing more efficient approximation strategies that balance accuracy with quantum resource requirements (such as circuit depth and ancilla usage) is of primary interest. Extending the framework to a broader class of nonlinearities while maintaining quantum compatibility and implementing the proposed approach on current and near-term quantum hardware will be explored, requiring modifications to the algorithmic structure.


\section*{CRediT authorship contribution statement}

Uditnarayan Kouskiya: Conceptualization, Methodology, Software, Formal analysis, 
Investigation, Data curation, Visualization, Writing – original draft, 
Writing – review \& editing. 
Caglar Oskay: Conceptualization, Supervision, Methodology, Formal analysis, Writing – review \& editing, 
Project administration, Funding acquisition.

\section*{Data availability}

Data supporting the findings of this study are available from the corresponding author upon reasonable request.

\section*{Declaration of competing interest}

The authors declare that they have no known competing financial interests 
or personal relationships that could have appeared to influence the work 
reported in this manuscript.

\section*{Acknowledgments}
The authors gratefully acknowledge the funding support from the National Science Foundation, CMMI Division, Mechanics
of Materials and Structures Program, United States of America (Award No: 2222404 and No: 2527249).

\appendix
\section{Solution to the simplified Neo-Hookean model}
\label{app:energy_func}

\subsection{Existence and Uniqueness of the Minimizer} \label{Sec:uniqueness}

Consider the energy functional introduced in Section~\ref{sec:NeoHookean_developement}:
\begin{equation}
\label{eq:energy_app}
\mathcal{E}[u]
=
\int_{\Omega_0}
\left[
\mu\left(u' + \frac{1}{2}(u')^2 - \ln(1+u')\right)
- B\,u
\right] \,\mathrm{d}X
-
\sum_{X\in\Gamma_N} \bar{P}\,u(X).
\end{equation}
with $u(X) = \bar{u}$ for $X\in \Gamma_D$. For  
\begin{equation*}
\mathcal{V}_0 := \left\{ v \in H^1(\Omega_0) \,:\, v = 0 \ \text{on } \Gamma_D \right\},
\end{equation*}
 the first variation of $\mathcal{E}$ in the direction $v\in\mathcal{V}_0$ is given by
\begin{equation}
\label{eq:first_variation_app}
\delta \mathcal{E}[u;v]
=
\lim_{\epsilon\to 0}
\frac{\mathcal{E}[u+\epsilon v] - \mathcal{E}[u]}{\epsilon}.
\end{equation}
A direct computation yields
\begin{equation}
\delta \mathcal{E}[u;v]
=
\int_{\Omega_0}
P(\lambda)\, v'\,\mathrm{d}X
-
\int_{\Omega_0} B(X)\,v\,\mathrm{d}X
-
\sum_{X\in\Gamma_N} \bar{P}\,v(X),
\end{equation}
where
\begin{equation}
    \label{eq:app_constitutive1d}
\lambda = 1 + u',
\qquad
P(\lambda) = \mu\left(\lambda - \lambda^{-1}\right).
\end{equation}
Thus, the condition
\begin{equation}
\label{eq:weak_criteria}
\delta \mathcal{E}[u;v] = 0 \quad \forall v \in \mathcal{V}_0
\end{equation}
is equivalent to the weak form of the governing equations presented in Section~\ref{sec:math_model}. Under sufficient smoothness, the last equation implies that $u$ satisfies the strong form. Consider the second variation of $\mathcal{E}$, in the direction $w \in \mathcal{V}_0$:
\begin{equation}
\label{eq:second_variation_app}
\delta^2 \mathcal{E}[u;v,w]
=
\int_{\Omega_0}
\mu
\left(
1 + (1+u')^{-2}
\right)
v'(X)\,w'(X)
\,\mathrm{d}X.
\end{equation}
By setting $w = v$, we obtain
\begin{equation*}
\delta^2 \mathcal{E}[u;v,v]
=
\int_{\Omega_0}
\mu
\left(
1 + (1+u')^{-2}
\right)
\big(v'(X)\big)^2
\,\mathrm{d}X.
\end{equation*}
Since physical admissibility requires $\lambda = 1+u' > 0$, we obtain
$$
1 + (1+u')^{-2} > 0.
$$
Hence, for $\mu > 0$, the integrand in \eqref{eq:second_variation_app} is strictly positive for any $v \neq 0$, which implies
$$
\delta^2 \mathcal{E}[u;v,v] > 0
\quad \forall v \in \mathcal{V}_0,\; v \neq 0.
$$
Therefore, the second variation defines a positive definite bilinear form on $\mathcal{V}_0$, and the energy functional $\mathcal{E}$ is strictly convex over the admissible space. Consequently, the minimizer of $\mathcal{E}$ is unique and also represents the solution of the weak form \eqref{eq:weak_criteria}.

\subsection{Exact Solution of the Strong Form} \label{app:exact_sol_deriv}
Based on the examples considered in this work, we consider a Dirichlet condition at $X=0$ and a Neumann (traction) condition at $X=L$, i.e.,
$$
\Gamma_D = \{0\}, \qquad \Gamma_N = \{L\}.
$$
The strong form of the governing equation is
\begin{equation}
\frac{\mathrm{d}}{\mathrm{d}X} P(\lambda(X)) + B(X) = 0,
\label{eq:strong_form_app}
\end{equation}
Integrating \eqref{eq:strong_form_app} from $X$ to $L$, and using the boundary condition $P(\lambda(L)) = \bar{P}$, we obtain
\begin{equation}
\label{eq:first_integral_app}
P(\lambda(X))
=
\int_{X}^{L} B(s)\,\mathrm{d}s + \bar{P}.
\end{equation}
Substituting the constitutive relation from \eqref{eq:app_constitutive1d} gives
$$
\mu\left(\lambda(X) - \frac{1}{\lambda(X)}\right)
=
\int_{X}^{L} B(s)\,\mathrm{d}s + \bar{P}.
$$
Define
$$
H(X)
:=
\frac{1}{\mu}\int_{X}^{L} B(s)\,\mathrm{d}s + \frac{\bar{P}}{\mu}.
$$
Then the governing relation becomes
\begin{equation}
\lambda - \frac{1}{\lambda} = H.
\label{eq:lambda_eq}
\end{equation}
Solving \eqref{eq:lambda_eq} for $\lambda$ yields
\begin{equation}
\lambda(X)
=
\frac{1}{2}
\left(
H(X) \pm \sqrt{H(X)^2 + 4}
\right).
\label{eq:lambda_analy}
\end{equation}
The physically admissible solution corresponds to $\lambda>0$, hence
\begin{equation}
u'(X) = \lambda(X) - 1,
\label{eq:strain_to_disp}
\end{equation}
where we only consider the positive branch of \eqref{eq:lambda_analy}, respecting the local material orientation.
Integrating and applying the Dirichlet boundary condition $u(0)=\bar{u}$, we obtain
\begin{equation}
u(X)
=
\bar{u}
+
\int_{0}^{X}
\left[
\frac{1}{2}
\left(
H(s) + \sqrt{H(s)^2 + 4}
\right)
- 1
\right]\,\mathrm{d}s.
\end{equation}

\subsection{Convergence of the Taylor and IHT Approximations} \label{app:constitutive_error}

The Taylor series expansion converges absolutely for $-1<u'<1$, which corresponds to $0<\lambda<2$ via \eqref{eq:strain_to_disp}. This translates to the requirement that body forces and the applied traction boundary condition must not be extreme enough to cause the resulting $u'$ to violate its allowed limit. Using \eqref{eq:lambda_analy}, we note that $\lambda>0$ gets automatically satisfied, while 
\begin{equation}
    \lambda<2 \quad\Longleftrightarrow\quad
\sup_{X \in [0,L]} H(X) < \frac{3}{2} \quad\Longleftrightarrow\quad  \left(\bar P + \sup_{X\in[0,L]}\int_X^L B(s)\,\mathrm{d}s\right)\;<\;\frac{3}{2}\mu .
\end{equation} 

For $0\leq u'\leq1$, the Taylor expansion forms an alternating series with terms decreasing in magnitude, and its truncation error obeys the standard alternating-series bound \cite{abbott2015understanding}. For $-1<u'<0$, the remainder can instead be bounded using the corresponding geometric tail:
\begin{equation}
\left|
\ln(1+u')
-
\sum_{i=1}^{N}\frac{(-1)^{i+1}}{i}(u')^i
\right|
\leq
\begin{cases}
\dfrac{|u'|^{N+1}}{N+1}, & 0\leq u'\leq 1, \\[2mm]
\dfrac{|u'|^{N+1}}{(N+1)(1-|u'|)}, & -1<u'<0.
\end{cases}
\end{equation}
For $|u'|<1$, the Taylor series converges absolutely, and the truncation error
decays geometrically, $O\!\left(|u'|^{N}\right)$. At $u'=1$ the series
converges only conditionally, and the bound weakens to $O\!\left(1/N\right)$, resulting in substantially slower convergence.
The IHT expansion \eqref{eq:IHT_approximation} with $N$ number of terms is the Taylor series of
$\ln\!\left(\tfrac{1+x}{1-x}\right)/2$ evaluated at $x=u'/(u'+2)$, so its truncation error can be bounded by a geometric tail,
\begin{equation}
    \left| \ln(1+u') - 2\sum_{i=1}^{N} \frac{1}{2i-1}\left(\frac{u'}{u'+2}\right)^{2i-1} \right|
    \le
    \frac{2}{1-x^2}\, \frac{|x|^{2N+1}}{2N+1},
    \qquad x = \frac{u'}{u'+2},
\end{equation}
valid for all $u'>-1$, since $|x|<1$ throughout this range, giving a
truncation error that decays geometrically as $O\!\left(|x|^{2N}\right)$.

The effect of these constitutive-approximation errors on the computed 
displacement field can be quantified through the corresponding stress errors. 
Let $\tilde{P}_N^T$ and $\tilde{P}_N^H$ denote the stress functions obtained 
by differentiating the strain energy densities corresponding to the Taylor 
and IHT approximations, respectively, with respect to $\lambda=1+u'$. 
The respective stress errors can be established as
\begin{align}
\varepsilon_T(u',N)
&:=
\left|P(u')-\tilde{P}_N^{T}(u')\right|
=
\frac{\mu |u'|^N}{1+u'},
\label{eq:taylor_stress_error}
\\
\varepsilon_H(u',N)
&:=
\left|P(u')-\tilde{P}_N^{H}(u')\right|
=
\frac{\mu}{1+u'}
\left|
\frac{u'}{u'+2}
\right|^{2N}.
\label{eq:iht_stress_error}
\end{align}

Let $u^*$ and $\tilde{u}_N$ denote the solutions obtained using the exact 
and an approximate constitutive law, respectively. From the strong  monotonicity 
of $P$ established in Appendix~\ref{Sec:uniqueness}, we have
$P'(\lambda)=\mu(1+\lambda^{-2})\geq\mu$. Thus, applying the mean value 
theorem to the exact stress function, and noting that the exact and approximate 
solutions correspond to the same equilibrium loading, the error in the 
displacement gradient can be bounded in terms of the corresponding constitutive 
stress error as
\begin{equation}
\left|\tilde{u}_N'(X)-u'^*(X)\right|
\leq
\frac{1}{\mu}\,
\varepsilon_{\bullet}\!\left(\tilde{u}_N'(X),N\right),
\end{equation}
where $\varepsilon_{\bullet}$, with $\bullet\in\{T,H\}$, denotes the 
corresponding Taylor or IHT-expansion-based stress error defined above. Integrating the displacement-gradient error along the bar and using the 
common Dirichlet condition 
$u^*(0)=\tilde{u}_N(0)=\bar{u}$ gives
\begin{equation}
\left|\tilde{u}_N(X)-u^*(X)\right|
\leq
\frac{1}{\mu}
\int_0^X
\varepsilon_{\bullet}\!\left(\tilde{u}_N'(s),N\right)
\,\mathrm{d}s.
\end{equation}
Consequently, provided the approximate displacement gradients remain bounded 
away from $u'=-1$, the displacement error inherits the geometric convergence 
of the corresponding stress approximation. For the Taylor approximation, 
this requires $\sup_X|\tilde{u}_N'(X)|<1$, whereas for the IHT approximation,
$\left|\tilde{u}_N'/(\tilde{u}_N'+2)\right|<1$ for all physically admissible
$\tilde{u}_N'>-1$.

Furthermore, the IHT expansion has been employed using a quadratic penalty in \eqref{eq:penaly_functional}. To quantify the error introduced by the finite penalty parameter $\mathcal P$, we consider one of the  optimality criteria of \eqref{eq:penaly_functional}, which leads to:
\begin{equation}
    \frac{\partial \mathcal{\tilde{E}}}{\partial y} = 0 \Rightarrow 2\mu y^2 - \mathcal{P} a^2 y + \big(2\mu + \mathcal{P} a u'\big) = 0, \qquad a := u' + 2.
\end{equation}
The last equation is a quadratic equation in $y$. Selecting the root that continuously approaches the exact constraint solution as $\mathcal P\to\infty$ gives 
\begin{equation}
    y_{\mathcal{P}}(u') = \frac{2\big(\mathcal{P} a u' + 2\mu\big)}{\mathcal{P} a^2 + \sqrt{\mathcal{P}^2 a^4 - 8\mu \mathcal{P} a u' - 16\mu^2}}, \qquad a = u'+2.
\end{equation}
 As $\mathcal{P}\to\infty$, $y_{\mathcal{P}}(u') \to y^*(u') := u'/a$, the exact solution of the constraint \eqref{eq:y_f(u)}, at the rate
\begin{equation}
    y_{\mathcal{P}}(u') - y^*(u') = \frac{2\mu\big(1+y^{*}(u')^2\big)}{\mathcal{P}\,a^2} + \mathcal{O}(\mathcal{P}^{-2}).
\end{equation}
The above result shows that the error 
introduced by the finite-penalty enforcement of the auxiliary constraint 
decays as $\mathcal{O}(\mathcal{P}^{-1})$. To examine how this error affects 
the corresponding energy, let $W^{\mathcal{P}}(u';\mathcal{P})$ denote the strain energy 
density obtained after minimizing the penalized formulation with respect to 
$y$:
\begin{equation}
    W^{\mathcal{P}}(u';\mathcal{P})
    :=
    \mu\left[
        u' + \frac{1}{2}(u')^2
        - 2\left(
            y_{\mathcal{P}} + \frac{1}{3}y_{\mathcal{P}}^3
        \right)
    \right]
    + \frac{\mathcal{P}}{2}
    \left(a y_{\mathcal{P}}-u'\right)^2.
\end{equation}
and let $W^N(u')$ denote the corresponding exact-constraint strain energy density obtained using the IHT expansion up to the cubic term:
\begin{equation}
    W^N(u')
    :=
    \mu\left[
        u' + \frac{1}{2}(u')^2
        -2\left(
            y^*(u')+\frac{1}{3}y^*(u')^3
        \right)
    \right],
    \qquad
    y^*(u')=\frac{u'}{u'+2}.
\end{equation} 
Substituting
$y_{\mathcal{P}}=y^*+\mathcal{O}(\mathcal{P}^{-1})$ gives
\begin{equation}
    W^{\mathcal{P}}(u';\mathcal{P})-W^N(u')
    =
    -\frac{2\mu^2\big(1+y^*(u')^2\big)^2}
    {\mathcal{P}(u'+2)^2}
    +\mathcal{O}(\mathcal{P}^{-2}).
\end{equation}
Thus, the penalty-induced error in the strain energy density also decays as
$\mathcal{O}(\mathcal{P}^{-1})$. 

Let $\tilde{\mathcal{E}}^{\mathcal{P}}$ and 
$\tilde{\mathcal{E}}^{N}$ denote the corresponding total potential-energy 
functionals obtained using the finite-penalty and exact-constraint strain 
energy densities, respectively. Since the above penalty-induced strain-energy-density error is 
$\mathcal{O}(\mathcal{P}^{-1})$ throughout the physically admissible domain, 
integration over $\Omega_0$ gives
\begin{equation}
    \tilde{\mathcal{E}}^{\mathcal{P}}[u]
    -
    \tilde{\mathcal{E}}^{N}[u]
    =
    \mathcal{O}(\mathcal{P}^{-1}).
\end{equation}
Hence, the finite-penalty functional approaches the corresponding 
exact-constraint IHT functional as $\mathcal{P}\to\infty$.

To relate this penalty error to the displacement solution, let
$u^{\mathcal{P}}$ and $u^N$ denote the minimizers of
$\tilde{\mathcal{E}}^{\mathcal{P}}$ and $\tilde{\mathcal{E}}^N$, respectively.
Let $P^N(u'):=\mathrm{d}W^N/\mathrm{d}u'$ denote the corresponding stress. It can be established that the truncated strain energy density remains strongly convex:
\begin{equation}
    \frac{\mathrm{d}P^{N}}{\mathrm{d}u'}
    =
    \frac{\mathrm{d}^2 W^{N}}{\mathrm{d}(u')^2}
    =
    \mu\left[
        1+
        \frac{16\big((u')^2+u'+2\big)}
        {(u'+2)^5}
    \right]
    > \mu,
    \qquad u'>-1,
\end{equation}
and hence the 
corresponding stress is strongly monotone, throughout the physically 
admissible range. Comparing the equilibrium conditions associated with
$\tilde{\mathcal{E}}^{\mathcal{P}}$ and $\tilde{\mathcal{E}}^N$ therefore 
bounds the displacement-gradient error by the penalty-induced stress 
perturbation, yielding
\begin{equation}
    \left\|u^{\mathcal{P}}-u^N\right\|_{H^1_0(\Omega_0)}
    =
    \mathcal{O}(\mathcal{P}^{-1}),
\end{equation}
where 
$\|v\|_{H^1_0}:=\|v'\|_{L^2(\Omega_0)}$ for the common Dirichlet condition 
at $X=0$. Consequently, in the present one-dimensional setting,
\begin{equation}
    \left\|u^{\mathcal{P}}-u^N\right\|_{L^\infty(\Omega_0)}
    \leq
    \sqrt{L}\,
    \left\|u^{\mathcal{P}}-u^N\right\|_{H^1_0(\Omega_0)}
    =
    \mathcal{O}(\mathcal{P}^{-1}).
\end{equation}

\section{Ansatz depth analysis} \label{sec:expressivity}
We investigate the ansatz depth requirements associated with polynomial approximations of the Neo-Hookean constitutive behavior as well as the problem size. In order to accelerate computations to quantify these relationships in a slightly broader range of polynomial order and problem size, the cost function computation is performed classically. The variational parameters generate a displacement profile through the ansatz, which is then used to evaluate the potential energy classically. The optimization algorithm in turn employs this potential energy computation for minimization.

We consider the problem from Section~\ref{sec:EX_P2} for our investigation with system sizes $n=3,\dots,7$ and polynomial orders of $N_{t_n}=3,4,5$ for the Taylor approximation.  For each $(N_{t_n},n,d)$ pair, 20 runs are performed, and the solution with the lowest $L_2$ error in the displacement field relative to a high-accuracy classical solution is recorded. The analyses are continued until ansatz depths where the number of control parameters $p$ equal (or just exceed) the number of degrees of freedom. The ansatz depths needed to achieve errors less than $10^{-3}$ and $10^{-7}$ (the smaller threshold is only achieved in the cases when $p\geq2^n$) are summarized in Table~\ref{tab:combined_threshold}. To achieve an error below $10^{-7}$, polynomial approximation order does not appear to affect the ansatz depth. We also observe achieving accurate solutions with relatively modest depths within the range of problem sizes considered.

\begin{figure}[htbp]
    \centering
    \begin{subfigure}[b]{0.48\linewidth}
        \centering
        \includegraphics[width=\linewidth]{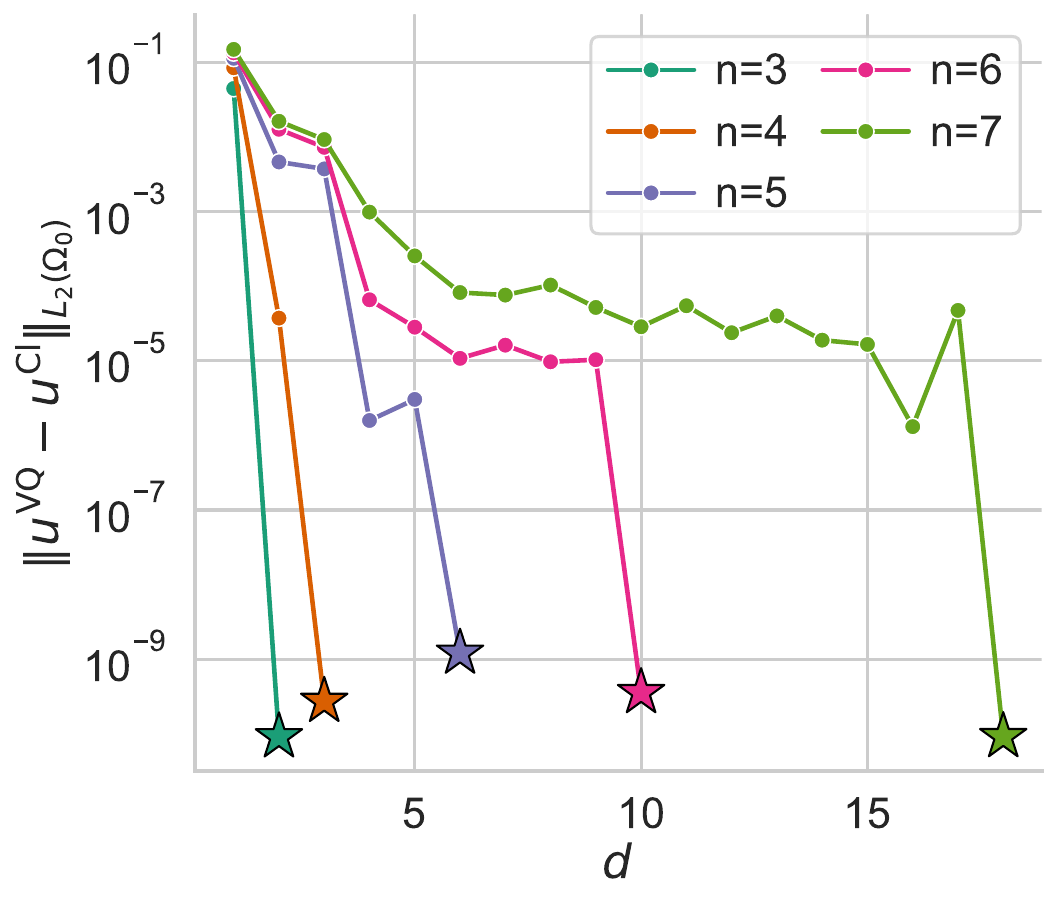}
        \caption{}
        \label{fig:error_vs_classical}
    \end{subfigure}
    \hfill
    \begin{subfigure}[b]{0.48\linewidth}
        \centering
        \includegraphics[width=\linewidth]{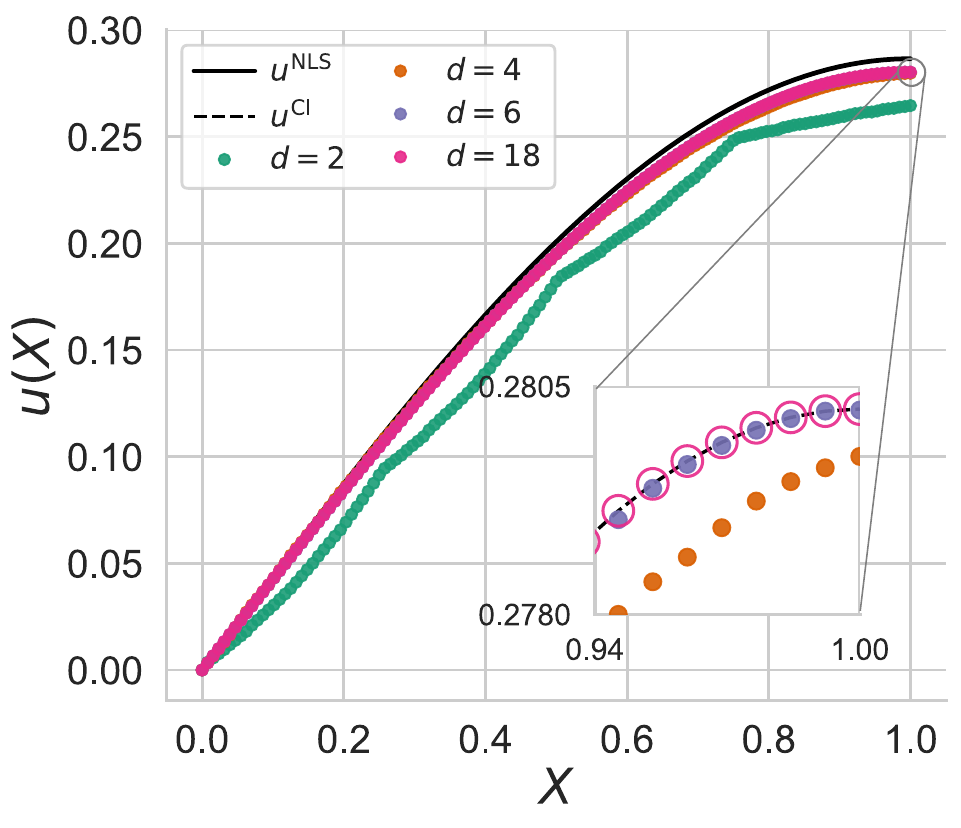}
        \caption{}
        \label{fig:error_vs_exact}
    \end{subfigure}

    \vspace{1em}

    \begin{subfigure}[b]{\linewidth}
        \centering
        \small
        \begin{tabular}{c*{10}{c}}
            \toprule
            \multirow{2}{*}{Order} & \multicolumn{10}{c}{$d$} \\
            \cmidrule(lr){2-11}
             & 1 & 2 & 3 & 4 & 5 & 6 & 7 & 8 & 9 & 10 \\
            \midrule
            3 & 0.0   & 20.0  & 40.0  & 65.0  & 75.0  & 90.0  & 95.0  & 95.0  & 95.0  & 95.0  \\
            4 & 100.0 & 100.0 & 100.0 & 100.0 & 100.0 & 100.0 & 100.0 & 100.0 & 100.0 & 100.0 \\
            5 & 0.0   & 35.0  & 45.0  & 80.0  & 75.0  & 70.0  & 65.0  & 85.0  & 90.0  & 80.0  \\
            \bottomrule
        \end{tabular}
        \caption{}
        \label{tab:vqa_success_n6}
    \end{subfigure}

    \caption{(a) Field $L_2$ error against the classical (non-VQA) Taylor-expansion-based energy functional up to 4th order, best of 20 independent trials per depth, plotted against the
    number of ansatz parameters $p$, for $n=3,\dots,7$; (b) Displacement profiles at $n=7$ for depths
    $d=2,4,6,18$, together with the true
    analytical solution $u^{\text{NLS}}$ and the classical minimizer $u^{\text{Cl}}$; (c) VQA
    success rate at $n=6$, shown
    as a percentage for truncation orders 3, 4, and 5 across depths
    $d=1,\dots,10$.}
    \label{fig:Expressivity}
\end{figure}

For $N_{t_n}=4$, the evolution of error as a function of ansatz depth with different problem sizes has been shown in Fig.~\ref{fig:error_vs_classical}.  Fig.~\ref{fig:error_vs_exact} shows the displacement profiles obtained for a few illustrative depths at $n=7$. We notice that the errors become negligible at considerably lower depths as compared to the depth required to exceed the problem size (starred points in Fig.~\ref{fig:error_vs_classical}). The success probabilities computed out of the 20 runs when problem size is set to $n=6$ for different polynomial orders are reported in table shown in Fig.~\ref{fig:Expressivity}, which shows higher success rates for larger depths.

\begin{table}[htbp]
    \centering
    \begin{tabular}{c cc cc cc}
        \toprule
        \multirow{2}{*}{$n$} & \multicolumn{2}{c}{$N_{t_n}=3$} & \multicolumn{2}{c}{$N_{t_n}=4$} & \multicolumn{2}{c}{$N_{t_n}=5$} \\
        \cmidrule(lr){2-3} \cmidrule(lr){4-5} \cmidrule(lr){6-7}
        & $10^{-7}$ & $10^{-3}$ & $10^{-7}$ & $10^{-3}$ & $10^{-7}$ & $10^{-3}$ \\
        \midrule
        3 & 2  & 2 & 2  & 2 & 2  & 2 \\
        4 & 3  & 2 & 3  & 2 & 3  & 2 \\
        5 & 6  & 4 & 6  & 4 & 6  & 4 \\
        6 & 10 & 4 & 10 & 4 & 10 & 4 \\
        7 & 18 & 6 & 18 & 5 & 18 & 6 \\
        \bottomrule
    \end{tabular}
    \caption{Minimum ansatz depths $d$ at which the  $L_2$ error in the displacement field, relative to a high-accuracy classical solution, first falls below the thresholds
    $10^{-7}$ and $10^{-3}$.}
    \label{tab:combined_threshold}
\end{table}

\section{Cost function  implementation for IHT expansion} \label{app:penalty_formulation}
For the functional~\eqref{eq:penaly_functional}, we employ a first-order
finite-element discretization for $u$ and approximate
$y$ using a piecewise constant field over the mesh. Specifically, we write
\begin{equation}
y(x) \approx \tilde y(x)
= \sum_{e=0}^{N_q-1} y_e \, \chi_e(x),
\end{equation}
where $\chi_e$ denotes the characteristic function of element $\Omega_e$. With these settings, the approximation of energy \eqref{eq:penaly_functional} based on a two-point Gauss quadrature can be given by:
\begin{align}
\label{eq:disc_penalty_energy}
\tilde{\mathcal{E}}(\boldsymbol{u},\boldsymbol{y})
&=
\sum_{e=0}^{N_q-1}
h_e
\left[ u'_e +
\frac{1}{2}\left(u'_e\right)^{2}
-2\!\left(y_e+\frac{1}{3}y_e^{3}\right)
+\frac{\mathcal{P}}{2}
\big(y_e(u'_e+2)-u'_e\big)^{2}
\right]
+ 
\tilde{\mathcal{E}}_B(\boldsymbol{u});
\\[6pt]
u'_e
&:=
\frac{u_{1_e}-u_{0_e}}{h_e}; \quad \tilde{\mathcal{E}}_B(\boldsymbol{u})
=
- \sum_{e=0}^{N_q-1}\sum_{g=0}^{1}
\frac{h_e}{2}\,
B\!\left(X^{(g)}\right)\,
\tilde u\!\left(X^{(g)}\right).\nonumber
\end{align}
The penalty term for any element can be expanded as:
\begin{equation*}
\frac{\mathcal{P}}{2}
\big(y_e(u'_e+2)-u'_e\big)^{2}
=
\frac{\mathcal{P}}{2}\Big(
y_e^2 (u'_e)^2
+ 4 y_e^2
+ (u'_e)^2
+ 4 y_e^2 u'_e
- 2 y_e (u'_e)^2
- 4 y_e u'_e
\Big).
\end{equation*}
Let $\bbu = (\bar{u},\bbv)$ be represented using control parameters as defined in \eqref{eq:quantum_rep_of_u}. Similarly, let $\bby\in\mathbb{R}^{N_q}$ be represented using:
\begin{equation*}
    \bby = \theta_0\,\widehat{V}(\bbtheta)\ket{0}.
\end{equation*}
A few terms in \eqref{eq:disc_penalty_energy} are treated in the main text in Section~\ref{sec:discrete_integral_circuits}. The remaining terms up to scaling factors, in the quantum-computing-implementable form are written below:
\begin{gather*}
\sum_{e=0}^{N_q-1}
h_e
 u'_e = u_{N_q} - u_0 = \lambda_0 \,
\left\langle \hat{v}\middle| \hat{m}^{(-1)} \right\rangle - \bar{u};
 \quad -2\sum_{e=0}^{N_q-1}
 h_e\,
 y_e = -2 \,\bbh^\top \,\bby = -2\, \theta_0 \| \bbh \| \left\langle \hat{h} \middle| \hat{y} \right\rangle; \\
-\frac23\sum_{e=0}^{N_q-1}
 h_e\,
 y_e^3 = -\frac23 \,\bby^\top \,(\bbh \odot \bby \odot \bby) = -\frac23\, \theta_0^3 \| \bbh \| \left\langle \hat{y} \middle|\,D_{\hat{h}} D_{\hat{y}} \middle| \hat{y} \right\rangle; \\
 2 \mathcal{P}\sum_{e=0}^{N_q-1}
 h_e\,
 y_e^2 = 2\mathcal{P} \,\bby^\top \,(\bbh \odot \bby) = 2\mathcal{P} \, \theta_0^2 \| \bbh \| \left\langle \hat{y} \middle|\,D_{\hat{h}} \middle| \hat{y} \right\rangle.
\end{gather*}
The term
\begin{align*}
\frac{\mathcal{P}}{2}\sum_{e=0}^{N_q-1} h_e\,(u'_e)^2 y_e^2
&=
\frac{\mathcal{P}}{2}
\underbrace{\left(\frac{\bar{u}^2 - 2 \bar{u}\, u_1 + u_1^2}{h_0}\; y_0^2\right)}_{\text{left-most element contribution}} + \,
T_{\text{bulk}},
\end{align*}
where 
\begin{gather*} u_1 = v_0 = \left\langle \hat{v}\middle| \hat{m}^{(1)} \right\rangle; \quad y_0 = \left\langle \hat{y}\middle| \hat{m}^{(1)} \right\rangle; \\ 
T_{\text{bulk}} = \lambda_0^2 \, \theta_0^2 \,\| \bbm^{(9)} \| \Big( \left\langle 
\hat{v}
\middle|
D_{\hat{y}}^2\,
D_{\hat{m}^{(9)}}
\middle|
\hat{v}
\right\rangle + \left\langle 
\widehat{A}^{\dagger} \hat{v}
\middle|
D_{\hat{y}}^2\,
D_{\hat{m}^{(9)}}
\middle|
\widehat{A}^{\dagger} \hat{v}
\right\rangle -2 \left\langle 
\hat{v}
\middle|
D_{\hat{y}}^2\,
D_{\hat{m}^{(9)}}
\widehat{A}^{\dagger} \middle|
\hat{v}
\right\rangle \Big)
\end{gather*}
and 
\begin{equation*}
m^{(9)}_i =
\begin{cases}
0, & i = 0, \\[4pt]
\dfrac{1}{h_{i}}, & i = 1,\ldots,N_q-1.
\end{cases}
\end{equation*}
Similarly, 
\begin{multline}
    - {\mathcal{P}}\sum_{e=0}^{N_q-1} h_e\,(u'_e)^2 y_e
=
-\mathcal{P}
\left(\frac{\bar{u}^2 - 2 \bar{u}\, u_1 + u_1^2}{h_0}\; y_0\right)  
\\ + \lambda_0^2 \, \theta_0 \,\| \bbm^{(9)} \| \Big( \left\langle 
\hat{v}
\middle|
D_{\hat{y}}\,
D_{\hat{m}^{(9)}}
\middle|
\hat{v}
\right\rangle + \left\langle 
\widehat{A}^{\dagger} \hat{v}
\middle|
D_{\hat{y}}\,
D_{\hat{m}^{(9)}}
\middle|
\widehat{A}^{\dagger} \hat{v}
\right\rangle -2 \, \left\langle 
\hat{v}
\middle|
D_{\hat{y}}\,
D_{\hat{m}^{(9)}}
\widehat{A}^\dagger \middle|
\hat{v}
\right\rangle \Big).
\end{multline}
We also have 
\begin{gather*}
  2 \mathcal{P} \sum_{e=0}^{N_q - 1} h_e y_e^2 u'_e  = - 2 \mathcal{P}  \left( \bar{u} \, y_0^2  - \lambda_0\,\theta_0^2 \,\left\langle 
\hat{y}
\middle|
D_{\hat{v}} \middle|
\hat{y}
\right\rangle + \lambda_0\,\theta_0^2 \| \bbm^{(10)} \|  \left\langle 
\hat{y}
\middle|
D_{\widehat{A}^\dagger \hat{v}}\,
D_{\hat{m}^{(10)}}
\middle|
\hat{y}
\right\rangle \right) \mbox{ and } \\
  -2 \mathcal{P} \sum_{e=0}^{N_q - 1} h_e y_e u'_e  =  2 \mathcal{P}  \left( \bar{u} \, y_0  - \lambda_0\,\theta_0 \,\left\langle 
\hat{y}
\middle|
\hat{v}
\right\rangle + \lambda_0\,\theta_0 \| \bbm^{(10)} \|  \left\langle 
0
\middle|
\widehat{V}^\dagger(\bblam)\,\widehat{A}\,
D_{\hat{m}^{(10)}}\,
 \widehat{V}(\bbtheta) \middle|
0
\right\rangle \right),
\end{gather*}
where $\bbm^{(10)} = [ 0, 1, 1\cdots 1]$.

\newpage
\printbibliography

\end{document}

%% file: temp-definitions.tex
\usepackage{amsmath}
\usepackage{amsbsy}
\usepackage{amssymb}
\usepackage{amscd}
\usepackage{amsfonts}

\newcommand{\beq}{\begin{equation}}
\newcommand{\eeq}{\end{equation}}
\newcommand{\beqs}{\begin{eqnarray}}
\newcommand{\eeqs}{\end{eqnarray}}
\newcommand{\beql}{\begin{equation} \label}

\newcommand{\p}{\partial}

\newcommand{\bbf}{\boldsymbol{f}}
\newcommand{\bbe}{\boldsymbol{e}}
\newcommand{\bbu}{\boldsymbol{u}}
\newcommand{\bbv}{\boldsymbol{v}}
\newcommand{\bby}{\boldsymbol{y}}
\newcommand{\bbp}{\boldsymbol{p}}
\newcommand{\bbq}{\boldsymbol{q}}

\newcommand{\bbm}{\boldsymbol{m}}
\newcommand{\bbll}{\boldsymbol{\ell}}
\newcommand{\bbA}{\boldsymbol{A}}
\newcommand{\bbX}{\boldsymbol{X}}
\newcommand{\bbP}{\boldsymbol{P}}
\newcommand{\bbB}{\boldsymbol{B}}

\newcommand{\bbF}{\boldsymbol{F}}
\newcommand{\bbN}{\boldsymbol{N}}
\newcommand{\bbI}{\boldsymbol{I}}
\newcommand{\bbT}{\boldsymbol{T}}
\newcommand{\bbh}{\boldsymbol{h}}

\newcommand{\bblam}{\boldsymbol{\lambda}}
\newcommand{\bbtheta}{\boldsymbol{\theta}}

\newcommand{\bbOmega}{\boldsymbol{\Omega}}

\newcommand{\dd}{\mathrm{d}}